\documentclass[11pt]{article}
\usepackage{arxiv}
\graphicspath{{Figures/}}
\newcommand{\hashid}[1]{{\ttfamily\footnotesize\seqsplit{#1}}}
\newcommand{\kms}{\ensuremath{\mathrm{km\,s^{-1}}}}

\title{Convergent representations of elastic-wave structure emerge across deliberately distinct seismic training routes}
\author[1]{Ziye Yu\thanks{Corresponding author: \href{mailto:yuziye@cea-igp.ac.cn}{yuziye@cea-igp.ac.cn}. ORCID: \href{https://orcid.org/0000-0002-1720-3811}{0000-0002-1720-3811}.}}
\author[4]{Yuqi Cai}
\author[2,3]{Xin Liu}
\affil[1]{Institute of Geophysics, China Earthquake Administration, Beijing 100081, China}
\affil[2]{Laboratory of Seismology and Physics of Earth's Interior, School of Earth and Space Sciences, University of Science and Technology of China, Hefei 230026, China}
\affil[3]{Institute of Advanced Technology, University of Science and Technology of China, Hefei 230088, China}
\affil[4]{University of Chinese Academy of Sciences, Beijing 100029, China}
\date{}
\hypersetup{pdftitle={Convergent representations of elastic-wave structure emerge across deliberately distinct seismic training routes},pdfauthor={Ziye Yu, Yuqi Cai, Xin Liu}}

\begin{document}

\maketitle

\begin{abstract}
Elastic-wave  propagation connects laboratory fracture with earthquakes, raising the prospect of seismic models that reuse waveform structure across physical scales. Here we  show that the phase picker PNSN  and the multi-task model SeismicXM  organize unseen laboratory acoustic emissions  around common waveform relations. Their representations agree on 512 held-out  records relative to matched random networks. Four earthquake-model families also track displaced laboratory arrivals without  fitting target neural weights.  Receiver-function classification is accessible to frozen pretrained features, although standardized random features remain competitive. Noise correlations test the extension to continuous propagation outputs: a nonlinear SeismicXM ensemble improves  on random topology,  while a period-wise median remains  more accurate. Waveform interventions and synthetic controls reveal that the observed agreement depends on feature calibration and the waveforms being compared. These results establish convergent waveform organization across distinct seismic training routes and  show that cross-scale arrival transfer is already accessible to compact models.
\end{abstract}

\section*{Introduction}

Seismology observes elastic waves across a vast range of source scales and processing pathways. Earthquake seismograms and laboratory acoustic emissions  (AEs) record rupture-generated waves at very different scales~\cite{ShiEtAl2024Labquakes,SheehanEtAl2025EQTtoAE,MastellaEtAl2026AEsNet}.  The source, medium and recording bandwidth  change, but recognizable arrivals and wave packets recur.  This recurrence raises a central question for seismic learning: can different models recover common waveform structure that remains useful beyond their training task?

Receiver functions (RFs) and ambient-noise  correlation functions (NCFs) extend this question to processed observations of Earth structure. RF deconvolution suppresses the common source signature and emphasizes near-receiver converted and reverberated phases~\cite{Phinney1964,Vinnik1977,Langston1979,AmmonEtAl1990,Ammon1991}.  Under suitable illumination and averaging conditions, noise correlation estimates an interreceiver response related to the elastic Green's function \cite{Wapenaar2004Green,ShapiroCampillo2004}. Ocean-wave interactions are an important source of ambient seismic noise, extending the range of excitation processes beyond rupture~\cite{ArdhuinEtAl2011Noise}. RFs and NCFs therefore complement rupture records by exposing propagation through different signal-processing operations.

 Deep models now perform a wide range of earthquake tasks~\cite{ZhuBeroza2019,MousaviEtAl2020EQT,SiEtAl2024,LiEtAl2024,ShengEtAl2025,CaiEtAl2026SeismicXM}.  Work on representational convergence asks whether differently trained models build similar internal descriptions of their inputs~\cite{HuhEtAl2024Platonic}. In seismology, this becomes a concrete question: do models judge the same unseen waveforms to be similar? Random networks provide a reference for useful structure supplied by architecture before source training~\cite{RahimiRecht2007,JacotEtAl2018,LuEtAl2022Frozen}. Comparing sample geometry then measures how learned waveform relations agree across models~\cite{KriegeskorteEtAl2008RSA,KornblithEtAl2019CKA,WilliamsEtAl2021Shape}. Representational agreement can vary with the population used  for comparison~\cite{CiernikEtAl2025Consistency}. Waveform interventions then probe which signal properties sustain that agreement.

We examine two deliberately distinct routes through the shared domain of earthquake arrivals. PNSN is a compact convolutional--recurrent model trained for regional phase picking \cite{CaiEtAl2025PNSN}. SeismicXM combines phase, polarity, event-type and waveform-reconstruction objectives in a convolutional--Transformer model~\cite{CaiEtAl2026SeismicXM}.  Their contrasting designs make common organization of new waveforms the central object of comparison. We use \emph{elastic-wave basis}  to refer to reusable relations among  observed waveform shapes.

 We bring together representation measurements and transfer tests across three seismic observations. Held-out AE waveforms anchor the primary PNSN--SeismicXM geometry comparison, with random networks, waveform interventions and  synthetic signals supplying complementary controls. Native outputs from PNSN, SeismicXM, PhaseNet and EQTransformer  test arrival transfer across the earthquake-to-laboratory scale change. RF classification tests transfer after deconvolution, while NCF dispersion probes nonlinear decoding into continuous propagation quantities~\cite{ShapiroCampillo2004,BensenEtAl2007}. Together, these experiments examine how earthquake-trained waveform representations extend from local arrivals to new physical outputs.

\section*{Results}

\subsection*{Distinct training routes retain transferable waveform structure}

\begin{figure}[!htbp]
\centering
\includegraphics[width=\textwidth]{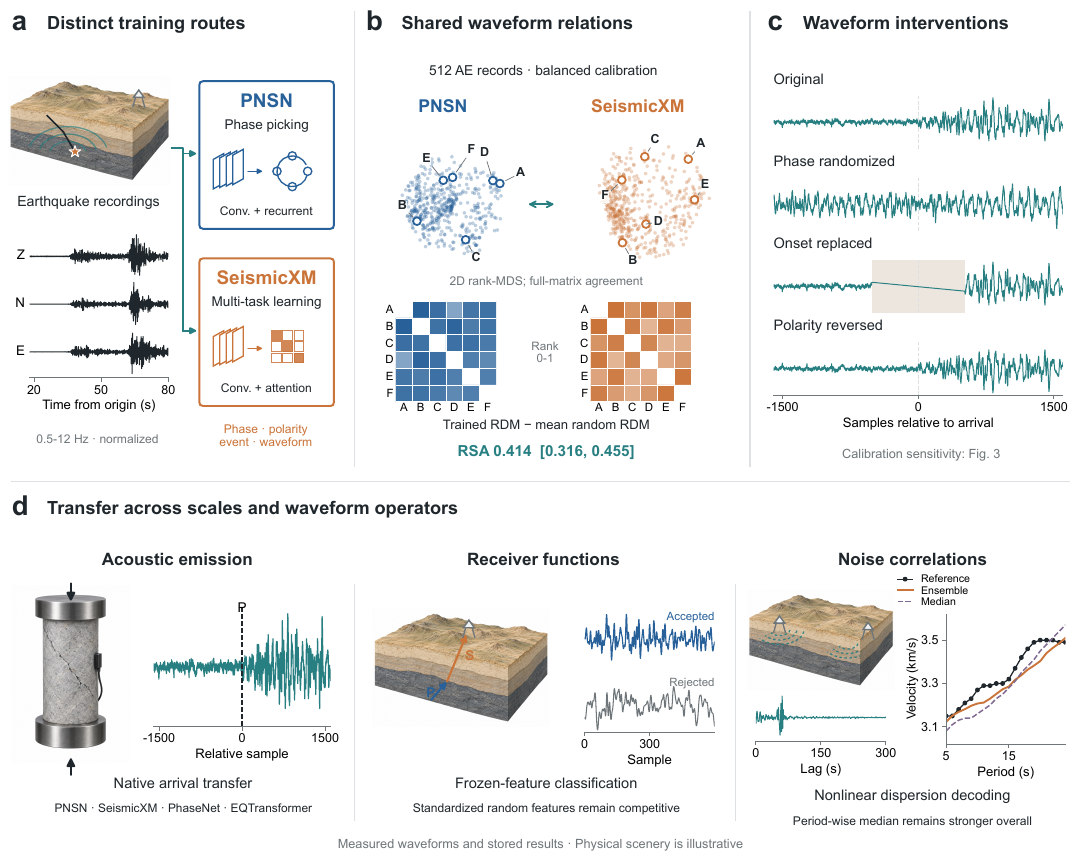}
\caption{\textbf{Distinct seismic training routes organize observed waveforms and support complementary transfer tests.} \textbf{a}, Recorded earthquake components introduce focused PNSN phase picking and broader SeismicXM multi-task training. Traces are filtered at 0.5--12 Hz and independently peak-normalized; network icons are schematic. \textbf{b}, Separate rank-MDS views of learning-delta cosine dissimilarities for the same 512 held-out positive AE records, after subtracting each topology’s mean of five random controls. Six matched records (A--F) connect the views to their exact full-population pairwise ranks in the lower matrices. No cross-model coordinate alignment is applied. MDS is a lossy display; the RSA and 95\% acquisition-channel bootstrap interval use the full matrices under the original balanced training calibration. Calibration sensitivity and other controls are reported in Fig. 3 and Supplementary Tables 18--20. \textbf{c}, A measured AE record, selected by median demeaned RMS, illustrates fixed-spectrum phase randomization, onset replacement and polarity reversal. Traces are demeaned and peak-normalized for display; zero marks the annotated arrival and shading marks the replacement window. \textbf{d}, Independent probes show the same AE record, accepted and rejected held-out RFs, and a station-disjoint NCF with its matching reference dispersion, SeismicXM ensemble prediction and training-period median. RF labels denote dataset quality, not predictions. RF and NCF examples are selected by within-class median variability and median test-curve error, respectively, for illustration. The period-wise median remains stronger over the full NCF test population. Four-family AE transfer does not establish four-model internal convergence. Rock and terrain backgrounds are generated illustrations, not measured scenes; stations, wavefronts and rays are schematic vector annotations. All empirical traces, geometry and dispersion curves are vector graphics.}
\label{fig:design}
\end{figure}

PNSN and SeismicXM retained useful wave structure after a change in waveform construction and a change in physical scale (Fig.~\ref{fig:design}). Frozen representations supported RF quality classification after deconvolution, and their released earthquake outputs detected and localized laboratory AE. These complementary tests ask whether training makes local morphology accessible outside its source task. NCF supplied a separate mechanism test of nonlinear continuous decoding, asking how those relations can be composed into period-indexed velocity.

The locked RF experiment establishes useful operator transfer for both routes: an affine classifier on each frozen pretrained encoder exceeded its corresponding random-encoder condition (Supplementary Table~3). That advantage was not invariant to the readout. A matched factorial applied balanced logistic regression to mean--maximum and first-position features, with and without training-only standardization. Across both interfaces, standardization brought random-feature performance close to the pretrained representations; none of the standardized pretrained gains had a waveform-bootstrap interval excluding zero (Supplementary Table~16; Supplementary Fig.~3a). RF therefore demonstrates accessibility through compact frozen features, not a general necessity for source pretraining. Target splits were fixed, regularization was selected on validation, and every tested protocol is reported.

PNSN makes compact sufficiency concrete: its retained encoder--recurrent path contains 345,600 parameters, yet supports both deconvolved RF morphology and native AE arrival transfer. This is a statement about what the released compact model can support, not a claim that its source supervision is uniquely responsible for RF classification. It motivates a separate question: whether the trained routes organize the same unseen waveforms in a similar way. Figure~\ref{fig:design} connects that test to waveform interventions and independent transfer probes. Complete parameter counts and label-efficiency controls are reported in Supplementary Tables~2 and 8.

\FloatBarrier

\subsection*{Native earthquake phase outputs transfer directly to laboratory acoustic emission}

\begin{figure}[!htbp]
\centering
\includegraphics[width=\textwidth]{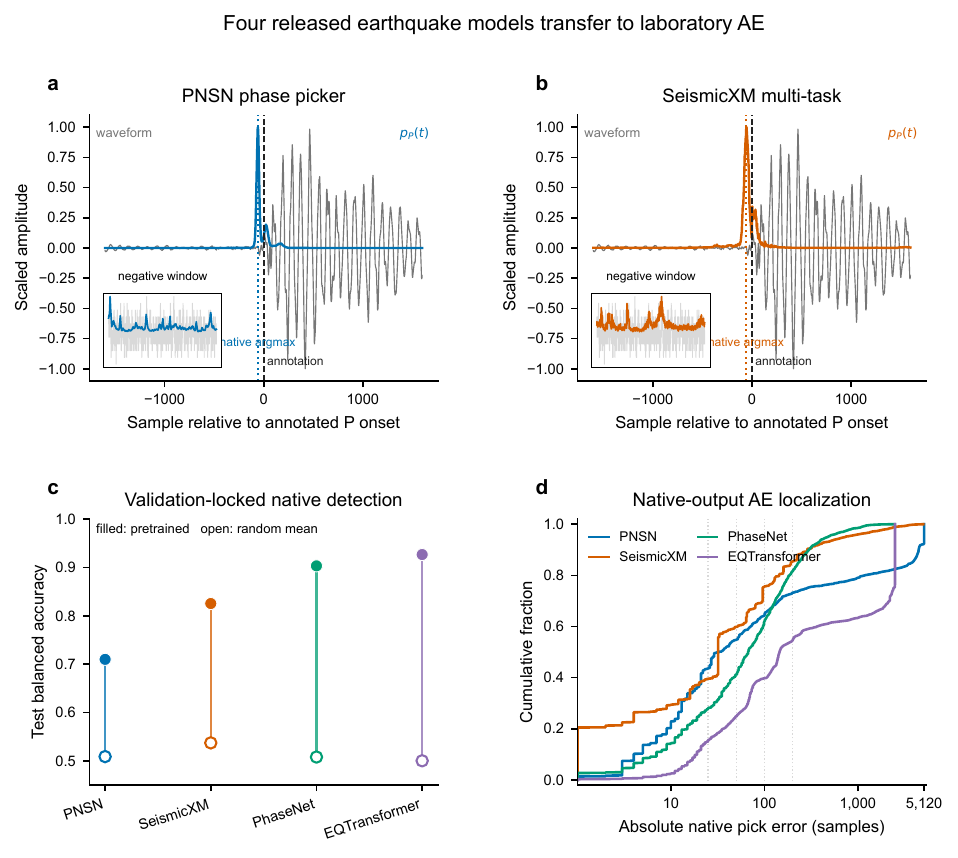}
\caption{\textbf{Four released earthquake models detect and localize laboratory AE.} \textbf{a,b}, Mechanistic examples from the primary PNSN and SeismicXM routes: the same positive AE waveform with their P-family curves; insets show the corresponding negative pretrigger window. Dashed and dotted lines mark the annotation and native argmax. \textbf{c}, Validation-locked test balanced accuracy for PNSN, SeismicXM, PhaseNet and EQTransformer, with each pretrained checkpoint paired to the mean of five complete random topologies. \textbf{d}, Pretrained native-output absolute pick-error distributions for all positive windows. PNSN and SeismicXM used the common 10,240-sample primary window; PhaseNet and EQTransformer used the common 6,000-sample generality window required by EQTransformer. These panels establish broader functional arrival transfer; the primary internal-representation convergence claim is quantified for PNSN--SeismicXM in Fig.~3, with the exploratory four-model matrix in Supplementary Fig.~2. Exact estimates and random-control localization distributions are in Supplementary Tables~3--5 and 12--13.}
\label{fig:nativeae}
\end{figure}

Earthquake-trained outputs remained sensitive to laboratory arrivals without learning a new neural decoder (Fig.~\ref{fig:nativeae}). We used the released P-family probabilities of PNSN and SeismicXM, the P output of PhaseNet, and the detection and P outputs of EQTransformer. Only scalar detection thresholds were selected on validation; all neural weights remained fixed. Arrival locations were read directly from the native P-probability maximum.

Both released phase heads detected AE above their matched random topologies, and their pretrained phase curves concentrated near annotated laboratory onsets (Fig.~\ref{fig:nativeae}a,b). PhaseNet and EQTransformer showed the same functional direction under a common shorter input accepted by both released models. Detection and localization improved relative to complete random instances in all four families (Fig.~\ref{fig:nativeae}c,d). Thus, arrival structure remained usable across the earthquake-to-laboratory scale change in models with substantially different internal designs.

The original annotation-centred crops permit a trivial constant-centre pick, so localization on those crops alone cannot establish arrival tracking. We therefore displaced annotations within common shorter windows while keeping all weights and centred-validation thresholds fixed. All four models outperformed the constant-centre reference in median localization error, with positive paired channel-bootstrap intervals for that contrast (Supplementary Table~17; Supplementary Fig.~3b). Prediction changes tracked annotation displacement, although EQTransformer retained substantial long tails and weaker displacement tracking. This supports transfer beyond a fixed window position, not deployment on unrestricted continuous streams. Complete original detection, tolerance and channel summaries are in Supplementary Tables~3--5 and 12--13; absent sampling-rate metadata restrict localization to samples.

\subsection*{Arrival-bearing waveforms support conditional representational convergence}

\begin{figure}[!htbp]
\centering
\includegraphics[width=\textwidth]{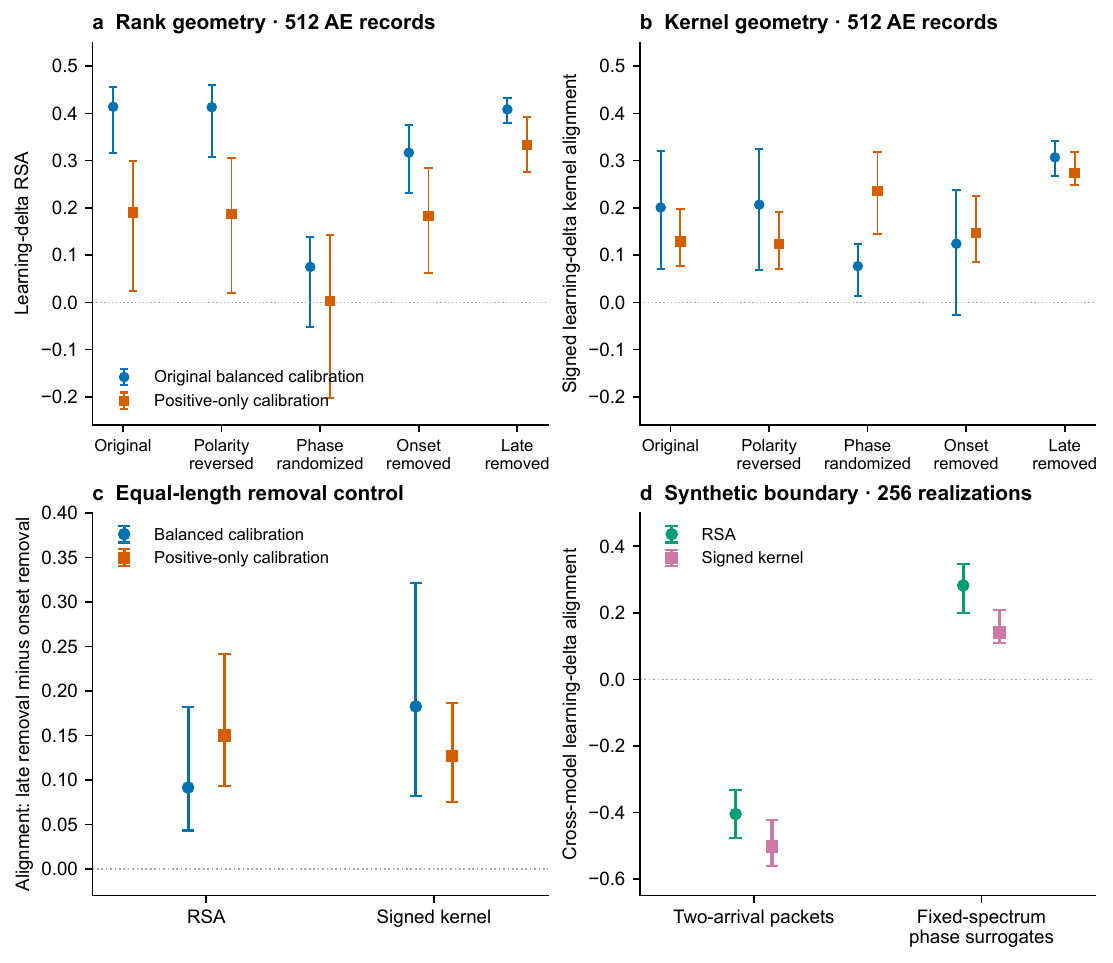}
\caption{\textbf{Expanded geometry separates reproducible agreement from conditional mechanism attribution.} \textbf{a,b}, Learning-delta RSA and signed kernel alignment on the same 512 positive AE records, using the original balanced training calibration or a positive-only training calibration. Five complete random topologies per route define the reference; bars are 95\% acquisition-channel bootstrap intervals from 500 shared resamples. Both calibrations retain positive original-input alignment, but intervention responses differ. \textbf{c}, Alignment after equal-length late-window replacement minus alignment after onset replacement; shared resamples preserve the pairing. Positive differences indicate relatively greater disruption at the onset, not its unique physical cause. \textbf{d}, The same models on 256 synthetic two-arrival scalar packets and fixed-spectrum phase surrogates, with 500 realization-bootstrap intervals. This boundary test does not simulate full elastodynamics. The original depth and random-control analysis is retained in Supplementary Fig.~4; the exploratory four-model matrix remains in Supplementary Fig.~2.}
\label{fig:convergence}
\end{figure}

PNSN and SeismicXM organized unseen onset-bearing AE waveforms around concordant relations relative to random topology (Fig.~\ref{fig:convergence}a,b). We compared which waveforms were similar within each model, allowing feature dimensions and coordinates to differ. Each pretrained representational dissimilarity matrix or centered kernel was referenced to the mean geometry of its complete random topology. This \emph{learning delta} describes a training-associated contrast, not a causal decomposition of source training; its signed kernel alignment is separate from conventional linear CKA.

For the original positive AE subset, both final-layer learning-delta metrics were stable under waveform resampling and separated from a shuffled-correspondence null (Supplementary Note~6 and Supplementary Fig.~4). An expanded evaluation retained positive alignment on 512 held-out positive records under both the original balanced training calibration and a positive-only calibration, with channel-bootstrap intervals above zero (Supplementary Table~18). The association therefore extends beyond the original small subset and is not simply a signal-versus-background separation. Its magnitude nevertheless changes with the feature origin and scale defined by the training population.

Interventions support a phase-sensitive interpretation under the original calibration, but qualify its generality. In the expanded sample, fixed-spectrum phase randomization reduced both alignment measures under the balanced calibration. Onset replacement reduced RSA, whereas its kernel contrast included zero. Equal-length replacement later in the waveform left greater alignment than onset replacement under both calibrations (Fig.~\ref{fig:convergence}c; Supplementary Table~19). With positive-only calibration, however, phase randomization increased kernel alignment and neither onset-versus-original contrast excluded zero. Polarity reversal produced little change under either calibration. Thus, relative sensitivity to waveform position is reproducible, but a calibration-independent joint phase/onset mechanism is not established.

Elementary envelope, spectrum and onset descriptors did not exhaust the positive-calibrated AE rank association, although this residualization is descriptive and cannot establish physical specificity. A separately generated family of scalar two-arrival packets provided a further boundary: the primary models' learning deltas were negatively aligned on the original packets, but positively aligned after phase randomization (Fig.~\ref{fig:convergence}d; Supplementary Table~20). This simplified signal family is not an elastic-wave solver. Its outcome nonetheless prevents extrapolating the observed AE convergence to an arbitrary collection of arrivals or to a universal physical representation.

The exploratory four-model matrix showed that this geometric agreement is selective (Supplementary Fig.~2 and Supplementary Table~14). On identical shorter AE crops, PNSN--SeismicXM and PNSN--EQTransformer aligned under both metrics. Comparisons involving PhaseNet aligned with the primary routes under RSA but not the signed kernel test; the other pairs did not separate from correspondence permutations. Functional arrival transfer is therefore broader than the evidence for internal geometric convergence. The primary PNSN--SeismicXM comparison supplies the combined depth and intervention evidence, while the four-model matrix establishes that successful transfer can coexist with distinct internal geometries.

\FloatBarrier

\subsection*{Continuous dispersion exposes the conditions for composing wave structure}

\begin{figure}[!htbp]
\centering
\includegraphics[width=\textwidth]{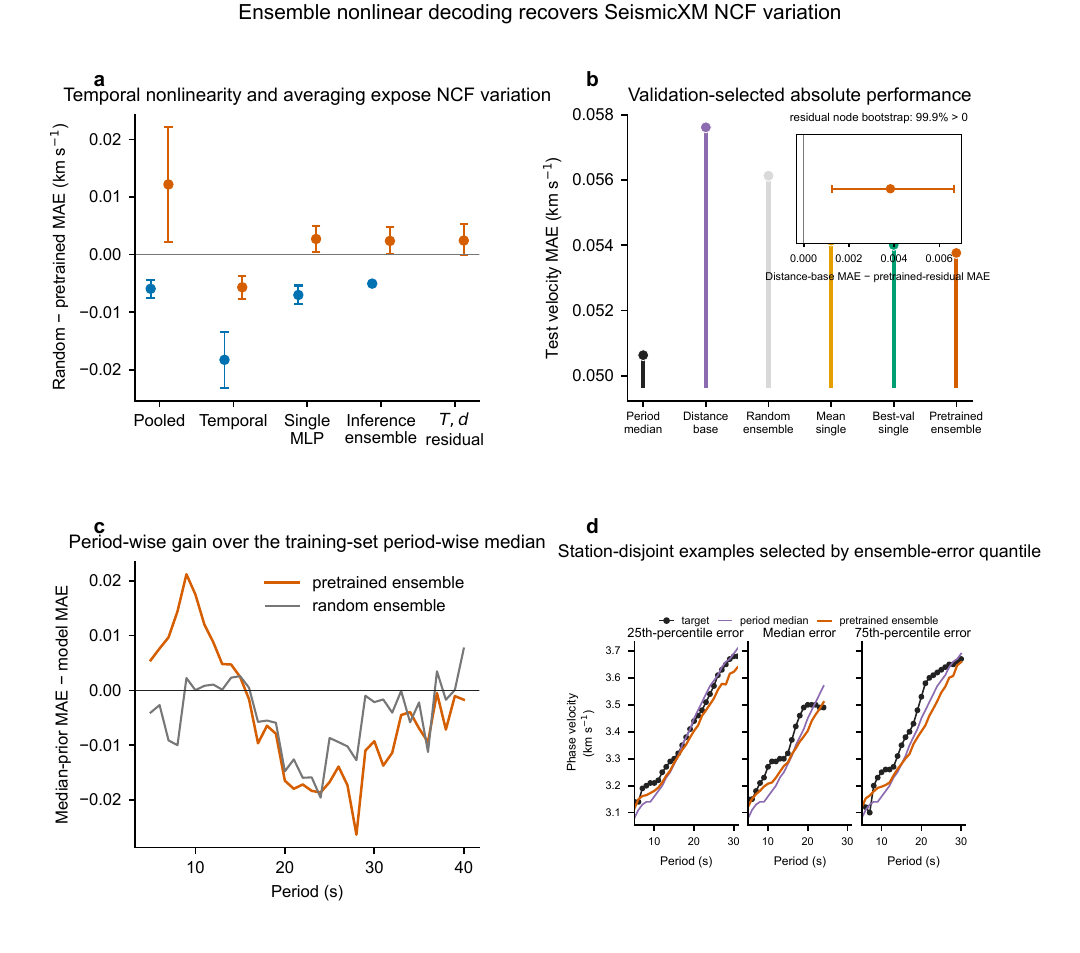}
\caption{\textbf{Ensemble nonlinear decoding recovers SeismicXM NCF variation.} \textbf{a}, Within-topology velocity-MAE gain across pooled, temporal-linear, single-MLP, inference-ensemble and period--distance residual readouts. Error bars show the corresponding seed or station-network-node uncertainty. \textbf{b}, Absolute station-disjoint performance of the marginal and distance baselines, random ensemble, individual members and pretrained ensemble; the inset shows improvement of the pretrained residual ensemble over the distance baseline. \textbf{c}, Period-wise gain over the training-set period-wise median. \textbf{d}, Held-out curves selected deterministically at representative pretrained-ensemble error quantiles for visualization only. All model and ensemble choices used validation data only.}
\label{fig:ncf}
\end{figure}

Mapping an NCF waveform onto a dispersion curve requires relations across propagation lag to be expressed as velocity across period. This makes NCF a mechanism test of nonlinear continuous decoding: it asks whether reusable waveform structure remains accessible when the output coordinates and source process change. For SeismicXM, a pooled affine readout showed a pretrained advantage at high absolute error. Preserving temporal position with an affine readout did not yield a corresponding gain. A nonlinear sequence readout exposed a lower-error pretrained advantage, which we evaluated with a validation-selected inference ensemble (Fig.~\ref{fig:ncf}).

The pretrained inference ensemble improved on matched random topology on station-disjoint paths, with a station-network bootstrap interval narrowly above zero (Supplementary Tables~6 and 7). Averaging modestly improved on individual members, and validity classification showed the same favorable direction. The training-set period-wise median nevertheless remained the stronger absolute velocity baseline. These comparisons answer distinct physical questions: the pretrained--random contrast detects a contribution from source learning within the tested waveform model, while the median tests whether that contribution is sufficient for competitive absolute dispersion prediction. Here the first condition was met and the second was not.

The waveform input also contains implicit geometry: its energy timing covaried strongly with interstation distance across every split (Supplementary Note~5). A residual ensemble improved on an explicit training-only period--distance baseline, establishing predictive information in the waveform beyond that relation. Its pretrained--random contrast favored source training but its interval included zero (Supplementary Tables~6 and 7). Thus, waveform information and a specifically pretrained contribution remain distinguishable even within continuous decoding.

The corresponding PNSN ensemble and supplementary architecture-native PhaseNet and EQTransformer readouts did not establish a favorable pretrained NCF advantage (Supplementary Tables~6 and 15; Supplementary Note~7). Arrival transfer therefore does not guarantee access to dispersion through a given interface. The staged decoder tests identify an accessible SeismicXM regime; a negative readout does not establish that a representation contains no relevant information. This boundary separates the broad transfer of arrival-sensitive outputs from access to continuous propagation quantities, which remains conditional on the model and decoder.

\FloatBarrier

\section*{Discussion}

Distinct seismic training routes preserve concordant relations among held-out AE waveforms despite differences in architecture and training objectives. PNSN--SeismicXM agreement persists under two training calibrations, while native arrival transfer extends across four model families. The distinction between shared waveform organization and task-specific prediction connects the convergence analysis to the wider problem of seismic transfer.

Common elastic-wave propagation provides a physical framework for interpreting reusable waveform morphology.  In a simplified linear point-source description, $u(t)=g(t)*s(t)$ convolves the source-time function $s(t)$ with the effective source--receiver response $g(t)$. Tensor components and instrument effects are suppressed in this notation. RF deconvolution and noise correlation expose different aspects of  the propagation response. Across these changes, arrivals and wave packets offer candidate relations for reuse, even when sources, scales and waveform construction differ. Earlier earthquake-to-AE studies established the practical reach of phase models~\cite{ShiEtAl2024Labquakes,SheehanEtAl2025EQTtoAE,MastellaEtAl2026AEsNet}. Here, that reach is examined alongside internal relational agreement and explicit tests of its limits.

Task-specific readouts make the distinction between waveform organization and prediction explicit. For waveform $x$, prediction is $\hat y_{m,\tau}=h_{m,\tau}(f_m(x))$, where $f_m$ denotes the route-specific encoder and $h_{m,\tau}$ the task-specific readout. Classification partitions the representation into decision regions, whereas regression assigns continuous values to it. The RF controls preclude claiming that learned physical structure is necessary for every successful transfer. Standardized random features remain competitive, even though the pretrained encoders support useful classification.

NCF tests access to propagation information in a different output coordinate. Under suitable illumination and averaging conditions, correlation waveforms approximate aspects of an interreceiver Green's-function response~\cite{Wapenaar2004Green}; dispersion curves describe velocity across periods~\cite{ShapiroCampillo2004,BensenEtAl2007}. The nonlinear SeismicXM ensemble exposes a small source-training advantage in this mapping, while the period-wise median remains the stronger absolute predictor. Residual prediction also reveals waveform information beyond the period--distance baseline, although its specific pretraining gain remains uncertain. The resulting contrast separates accessible waveform information from competitive absolute prediction. The comparison does not isolate source suppression as the cause of the NCF limit. RF processing also suppresses source signatures, while illumination, data shift and decoder design remain alternative influences.

The evidence establishes convergence conditional on the trained routes and evaluated waveforms, consistent with broader work on representational agreement and its dataset dependence~\cite{HuhEtAl2024Platonic,CiernikEtAl2025Consistency}. Intervention responses vary with calibration and metric, synthetic packets do not reproduce the AE alignment, and the four-model matrix is selective. Architecture, parameter count, source lineage and objectives co-vary between the primary routes, each represented by a single released checkpoint. Related earthquake lineage and arrival supervision are plausible shared ingredients, but identifying their contributions requires controlled source training. Routes without arrival supervision and non-seismic models trained on impulsive signals would test that attribution and elastic-wave specificity. A shared latent manifold and a causal role for the aligned component remain untested.

The population and decoder comparisons define the remaining scope of inference. RF transfer concerns unique archived waveforms, AE a later acquisition, and NCF station-disjoint paths within one regional system. Missing RF event/station and AE specimen identities limit broader population inference; waveform resampling does not quantify source-training seed variation. The additional controls are post-review sensitivity analyses, with post-hoc calibration checks, rather than independent external validation. NCF decoder stages also differ in capacity and optimization, so their comparison identifies a successful representation--readout combination rather than isolating nonlinearity alone.

Together, the results show that common waveform organization can emerge across distinct seismic training routes. The compact PNSN route already supports the tested RF classification and native AE transfer, making foundation-scale pretraining unnecessary for those outcomes. These findings motivate a design priority for seismic learning: preserve reusable waveform relations and evaluate their accessibility as source scale, waveform construction and target output change.

\section*{Methods}

\subsection*{Source encoders and frozen counterfactual}

The SeismicXM source checkpoint is the official general-purpose \path{seismicxm.middle.pt} linked by the upstream model zoo (repository commit \hashid{2d8077c62b6600e94d71a512b704b6fd6902f91d}; checkpoint SHA-256 \hashid{671d02d677c25c3d075963889602299ec71f52c724470f2fa85bb28035fe1528}). Six convolutional blocks feed a bidirectional positional encoder and three Transformer blocks with 1,024-dimensional tokens and eight heads. Source objectives include regional phase probabilities, P polarity, event type and waveform reconstruction~\cite{CaiEtAl2026SeismicXM}. The complete released model contains 51,895,545 parameters; the retained representation backbone contains 41,332,096.

PNSN v3 (repository commit \hashid{1f955f4d8cffbc5c43584f02ef14a0da2b0aeb6b}; checkpoint SHA-256 \hashid{9f626e5fff4e9390c88e43c2f6230802496163b5d6eefee05e1b6ac7ccebf9e8}) comprises seven convolution--batch-normalization--rectified-linear blocks, two bidirectional LSTMs and its released transposed-convolution phase decoder~\cite{CaiEtAl2025PNSN}. Its complete released model contains 450,485 parameters, of which 345,600 belong to the retained encoder--recurrent representation path. SeismicXM retained its released six-stage phase decoder. Counts sum each tensor returned by \texttt{named\_parameters()} once, exclude non-parameter buffers and exclude downstream target readouts from the retained-path values. Strict loading matched every checkpoint tensor in both complete models. Random controls reconstructed the same complete modules after setting the paired seed and loaded no checkpoint. RF, NCF and representation analyses extracted intermediate or final encoder sequences; native AE transfer kept the original phase decoder attached and unchanged.

The additional AE generality test used the original PhaseNet and conservative original EQTransformer checkpoints distributed through SeisBench~\cite{ZhuBeroza2019,MousaviEtAl2020EQT,WoollamEtAl2022}. Their complete released models contain 268,443 and 376,935 parameters, respectively. Five complete random instances of each exact topology supplied controls. PhaseNet used its native P output; EQTransformer used its detection and P outputs. Target traces were repeated into three identical channels for all four models, supplying identical scalar waveforms rather than reconstructed physical Z, N and E components.

\subsection*{Receiver-function data}

We used the public arrays from Gong \emph{et al.}~\cite{GongEtAl2022}, containing 5,456 radial RFs of 600 samples (4,614 rejected and 842 accepted before cleaning). Byte-level waveform hashing removed 801 exact duplicates; no duplicate group contained conflicting labels. The 4,655 unique traces comprised 3,944 rejected and 711 accepted examples. A locked stratified split contained 3,257 training, 699 validation and 699 test traces. Each was demeaned and divided by its maximum absolute amplitude. Row-level event, station, back-azimuth and ray-parameter identifiers were absent. The RF estimand is waveform-unique interpolation; station-, event- and structure-disjoint evaluation requires additional metadata.

\subsection*{Acoustic-emission data}

The AE development and later-acquisition HDF5 files from the cited deposition~\cite{Li2025AEbagging} contained 10,044 and 2,085 raw records from 12 channels. Hashing removed 346 development and 12 test duplicates, with no cross-split hash overlap. The chronologically last 10\% within each development channel formed validation, leaving 8,728 training and 970 validation records; the later file supplied 2,073 test records. Each record contained 40,000 samples and a P annotation at sample 20,000.

Native-output evaluation used a fixed 10,240-sample window, matching the released PNSN inference length and remaining exactly divisible by the SeismicXM encoder stride. The positive window placed the annotation at sample 5,120. The negative window ended 100 samples before the annotation. The same windows entered both models. PNSN inputs were demeaned and divided by per-channel standard deviation, following its released picker; SeismicXM inputs were demeaned and divided by maximum absolute amplitude, following its released wrapper. Both original decoders returned five-class softmax probabilities in the order Noise, Pg, Sg, Pn and Sn. P-family probability was the sum of Pg and Pn. Detection used its temporal maximum and picking used its argmax on the unchanged output grid, whose stride relative to input was one for both complete models. A single global scalar detection threshold maximized validation balanced accuracy with F1 and proximity-to-0.5 tie breaking, then remained locked on test. This was direct native-output transfer with no AE-trained neural weights: no parametric calibration, temperature parameter, target MLP, $1\times1$ head or AE-labelled weight entered the experiment.

Absolute pick errors were summarized across all positive windows and separately among positives whose native detection score exceeded the locked threshold. We reported the median, interquartile range, 90th percentile, complete empirical cumulative distribution and fractions within $\pm25$, $\pm50$, $\pm100$ and $\pm200$ samples. Matched-random distributions used the five complete random topologies. The polarity intervention multiplied every test waveform by $-1$ and reused the original validation threshold without refitting. Missing specimen, loading-stage, sampling-rate and synchronized multi-channel event identifiers define the result as later-acquisition, per-record transfer with channel-block uncertainty.

For PhaseNet and EQTransformer, a common 6,000-sample crop placed the positive annotation at sample 3,000 and ended the negative crop 100 samples before the annotation. PhaseNet inputs were demeaned and standardized per repeated component. EQTransformer inputs were demeaned, divided by their joint standard deviation and tapered at the edges. The central 5,000 outputs were scored after a 500-sample edge exclusion on each side. Threshold selection, test locking and matched-random evaluation otherwise followed the primary native-output protocol.

\subsection*{Ambient-noise correlation data and geographic split}

The public SeisDispFusion-NCF HDF5 file~\cite{Yu2026SeisDispFusionNCF} contained 44,359 waveform entries representing 41,600 nominal station pairs. Station endpoint order was repaired from source filenames. Nineteen station codes whose observed coordinates varied by more than $0.05^{\circ}$ were excluded, removing 4,213 pairs. The remaining 375 stations were standardized by longitude and latitude and partitioned by deterministic five-cluster $k$-means. The cluster with 132 stations and 3,025 within-cluster entries was held out (selected as the largest within-cluster population; five-cluster baseline shifts are reported in Supplementary Table~10). The other 243 stations supplied 15,755 training and 1,769 validation entries; 16,838 cross-boundary pairs were excluded. Test station overlap was zero, as were train/test pair, exact waveform and complete target-curve hash overlaps.

Each input was a one-component NCF on 1,501 samples from 0 to 300~s at 0.2~s intervals, demeaned and divided by its standard deviation. Targets were phase velocities at 2--50~s in 1-s increments plus a 49-element validity mask. Coordinates, distance and the explicit time array did not enter a neural model. Absolute sample position was retained, however, and thus encoded propagation-lag information. Energy-centroid and envelope-peak correlations with distance were computed directly from every waveform as a geometry analysis.

\subsection*{Target readouts and optimization}

The locked RF condition froze the encoder and trained an affine classifier on PNSN mean--maximum pooled features or on the first SeismicXM waveform token; each initialization comparison used the same architecture-specific interface. The native AE experiment used no target readout. The pooled NCF head mapped mean--maximum features to 49 velocity and 49 mask outputs. Within-topology gains isolate source training; absolute scores retain each released architecture and interface.

For NCF mechanism probes, parameter-free layer normalization was applied at each temporal position. A kernel-7 convolution with padding 3 produced 98 channels. The temporal-linear branch used a shared affine map over all 1,501 positions. The temporal-MLP branch used 1,501--256--64--1 layers with GELU activation and dropout 0.1, shared across the 98 velocity/mask outputs. The branches contained 67,456 and 704,064 trainable parameters (temporal linear) or 466,979 and 1,103,587 (temporal MLP) for PNSN and SeismicXM, respectively. These readouts form staged tests of temporal position and nonlinearity; every within-stage initialization comparison uses the same decoder and training budget. Head capacity and optimization settings differ across stages, so these contrasts identify successful representation--readout combinations without isolating a causal effect of nonlinearity alone.

All trained heads used AdamW with weight decay $10^{-4}$, gradient norm clipped at 5, and validation early stopping after five stale epochs. The head learning rate was $10^{-3}$ for pooled probes and $10^{-4}$ for temporal probes. RF used inverse-frequency cross-entropy. NCF minimized masked smooth-$L_1$ velocity loss plus 0.2 times mask binary cross-entropy and selected the lowest validation velocity MAE. Classification thresholds were selected on validation by balanced accuracy and then locked for test. Formal RF comparisons used ten paired seeds (20260826--20260835); the temporal-linear diagnostic used five and the single temporal MLP used ten.

\subsection*{Direct representational convergence}

RF and AE convergence used identical held-out waveforms in identical order for PNSN and SeismicXM. Balanced pools contained 192 training and 128 test examples per task; balanced sampling controlled class composition and made the high-temporal-resolution geometry calculation tractable. AE used the same 10,240-sample positive and pretrigger-negative construction as native transfer. We extracted the PNSN convolutional sequence and final bidirectional-LSTM sequence, and the SeismicXM positioned embedding and final waveform-token sequence. Adaptive temporal mean and maximum summaries used four bins for RF and 16 for AE. Each model, initialization and layer was standardized with its own training-subset mean and standard deviation; test samples never contributed scaling statistics.

For standard RSA, we formed cosine RDMs and correlated their upper triangles with Spearman's $\rho$\cite{KriegeskorteEtAl2008RSA}. Conventional linear CKA was computed only from unmodified total centered Gram matrices~\cite{KornblithEtAl2019CKA}. With $H=I-\mathbf{1}\mathbf{1}^{\mathsf T}/n$ and standardized descriptor matrix $X_M$, the kernel was $K_M=HX_MX_M^{\mathsf T}H$. Because total geometry includes relations already present at random initialization, the primary kernel estimand was
\begin{equation}
K_M^\Delta=K_M^{\mathrm{pretrained}}-\frac{1}{R}\sum_{r=1}^{R}K_{M,r}^{\mathrm{random}},
\end{equation}
and cross-model learning-delta kernel alignment was
\begin{equation}
A_\Delta=
\frac{\langle K_{\mathrm{PNSN}}^\Delta,K_{\mathrm{SeismicXM}}^\Delta\rangle_F}{\lVert K_{\mathrm{PNSN}}^\Delta\rVert_F\lVert K_{\mathrm{SeismicXM}}^\Delta\rVert_F}.
\end{equation}
$A_\Delta$ is a signed normalized kernel inner product and is not conventional non-negative CKA. Learning-delta RSA applied the equivalent subtraction to cosine RDMs before correlating their upper triangles.

For a random control realization $r$, its reference excluded itself:
\begin{equation}
K_{M,r}^{\Delta}=K_{M,r}^{\mathrm{random}}-
\frac{1}{R-1}\sum_{r'\ne r}K_{M,r'}^{\mathrm{random}},
\end{equation}
with the same leave-one-out rule for RDMs. Conditions were pretrained--pretrained, pretrained--random, random--pretrained and matched random--random. Waveform uncertainty used 500 stratified bootstrap resamples; stability used 250 repeated balanced 75\% subsets; the null independently permuted SeismicXM sample correspondence 1,000 times while preserving each model's internal ordering across its pretrained and random instances. Interventions were Fourier phase randomization at fixed amplitude, $x(t)\rightarrow-x(t)$ polarity reversal and replacement of the central $\pm512$-sample AE onset neighbourhood by a boundary-matched linear path. Random controls received the same transformed inputs.

The exploratory four-model matrix retained the same locked AE record selection but used common 6,000-sample crops and the positive test subset. PNSN used its final bidirectional recurrent sequence and SeismicXM its final waveform-token sequence, each summarized on a 16-bin adaptive mean--maximum grid. PhaseNet used its flattened encoder bottleneck; EQTransformer used the centre position of its shared post-attention sequence. Each model and initialization was standardized from its own training subset. Five complete random topologies defined each learning delta. Pairwise uncertainty used 1,000 waveform bootstrap resamples, and correspondence nulls used 1,000 independent permutations. This was an exploratory common-input sensitivity analysis and did not replace the primary PNSN--SeismicXM estimand on 10,240-sample windows.

\subsection*{Post-review robustness controls}

The extensions retained the official checkpoints and locked source-data splits. The RF factorial crossed mean--maximum versus first-position final features with raw versus training-standardized features, using one pretrained and five complete random instances per route. Balanced logistic regression selected $C\in\{0.01,0.1,1,10\}$ on validation; all protocols and class-stratified paired test uncertainty are reported. AE position tests paired centred and uniformly displaced annotations on 512 channel-stratified test records, with thresholds fitted only on 192 centred validation records. Common 6,000-sample content was right-padded to 6,144 for the primary models to preserve their exact native output grid; padded outputs were excluded, with no interpolation.

Expanded final-layer AE geometry used 512 positive test records, the original 16-bin descriptor and five random instances per route. An initial positive-only 192-record training calibration was followed by an explicitly post-hoc sensitivity test retaining the original balanced calibration on identical test features. Both are reported. Phase, polarity and onset interventions were paired with equal-length late-window removal. Uncertainty used 500 shared acquisition-channel resamples. Rank-RDM residualization controlled descriptive envelope, spectrum and onset statistics; it was not treated as a causal estimate. A synthetic scalar two-arrival Ricker-packet diagnostic varied delay, width, amplitude, origin and noise, with independent training/test realizations and fixed-spectrum phase surrogates. Its 500 realization resamples and known-delay associations test a specified signal family, not a full elastic-wave solution. Supplementary Note~9 gives the complete procedures, all outcomes and the distinction between planned extensions and post-hoc calibration sensitivity.

\subsection*{NCF inference ensemble and residual decoding}

The inference ensemble used ten independently initialized and optimized temporal-MLP members already selected by their own validation histories. Each member applied parameter-free channel normalization at every sequence position, a kernel-7 convolution producing 98 period-specific velocity/validity trajectories, fixed interpolation to 1,501 positions and a shared 1,501--256--64--1 GELU MLP. Velocity predictions and validity probabilities were averaged across members at inference. We predeclared $K\in\{5,10\}$ in seed order, selected $K$ by ensemble validation velocity MAE separately for each backbone and initialization, selected the validity threshold on the same validation split, and evaluated the test split once. Diversity was quantified by mean pairwise residual correlation and mean pairwise absolute prediction difference.

The distance-conditioned residual experiment first fitted $v_{\mathrm{base}}(T,d)$ on training data only. Polynomial distance degree and ridge regularization were selected on validation, locked and used in separate fits for each period. An architecture-matched SeismicXM temporal MLP predicted $\Delta v=0.5\tanh(z)$ and reconstructed $\hat v=v_{\mathrm{base}}(T,d)+\Delta v$, clipped to the declared 0.5--4.5~\kms range. Three independently optimized members were trained and averaged for each initialization, with the ensemble size fixed before test evaluation. Pretrained and random residual conditions shared seeds, data order, temporal-head capacity, optimizer, batch size and epoch budget; only source checkpoint loading differed.

\subsection*{Transparent NCF baselines}

The period-median baseline predicted the training-set median velocity independently at each valid period. Distance-only ridge regression used polynomial distance degrees 1--4 and $\alpha\in\{0,0.01,0.1,1,10,100\}$, selected on validation MAE. Waveform descriptors included envelope timing, moments, crest factor, zero crossings, spectral centroid and entropy, and 24-bin summaries of waveform, envelope and log spectrum. Standardized waveform-only and waveform-plus-distance ridge models used the same validation-selected regularization. A log-travel-time target was also tested before converting predictions back to velocity. The locked test region was evaluated once after selection. The period-median and distance-only procedures were repeated for all five geographic clusters to quantify regional shift; neural encoders were not retrained for the four additional clusters.

\subsection*{Uncertainty and physical diagnostics}

Seed results are reported as mean $\pm$ sample s.d. For target performance, we performed 10,000 bootstrap resamples. RF sampling was stratified within class over held-out waveforms after averaging paired seed contributions. AE used acquisition-channel block bootstrap; NCF used both station-pair sampling and a station-network-node bootstrap that reweighted all incident test pairs for sampled stations. The NCF procedure was applied pairwise to random-minus-pretrained and period-median-minus-pretrained inference-ensemble MAE, and to random-residual-minus-pretrained-residual and period--distance-baseline-minus-pretrained-residual MAE using the stored test predictions. Direct convergence used the separate waveform bootstrap, repeated balanced subsampling and correspondence permutation described above. Convergence values are conditional on the single released checkpoint per route; resampling quantifies held-out waveform uncertainty, not source-training seed variation. Percentile 95\% intervals are reported. No resampling scheme supplies unavailable RF event/station or AE specimen metadata.

For NCF, pointwise velocity errors were converted to relative velocity, travel time per 100~km and travel time over each actual path. First and second finite differences and adjacent-period trend signs measured curve geometry. These diagnostics do not constitute inversion. The favorable gain convention was pretrained minus random for higher-is-better metrics and random minus pretrained for errors.

\subsection*{Reproducibility and reporting}

The primary archive contains 240 formal readout runs, 20 temporal-linear runs, 40 independent temporal-MLP members, 12 complete-model native AE evaluations, direct convergence outputs for both layers and every intervention, the four-model common-input convergence matrix, and matched residual NCF runs. It includes 338 selected downstream checkpoints; all 79 checkpoint-dependent pretrained SeismicXM run records identify the official \path{seismicxm.middle.pt} route. The accompanying robustness extension additionally retains 48 selected affine RF heads, native displaced-window predictions, expanded and synthetic relational matrices, paired uncertainty arrays and code/checkpoint provenance. Cited waveform data and feature caches are excluded. Separate executable gates reproduce the original numerical archive~\cite{Yu2026ElasticWaveArchive} and the additional controls from stored artifacts; the latter also regenerates Supplementary Tables~16--20 exactly. Figure PDFs, 300-dpi PNGs and non-waveform CSV source data accompany the analysis code.

\section*{Data availability}

The receiver-function arrays are publicly downloadable from the repository associated with Gong \emph{et al.}~\cite{GongEtAl2022} at \url{https://github.com/gongchang96/Receiver-Function-Quality-Control}; that repository does not declare a data license. The AE records are held in the restricted-access \emph{AEbagging model} Zenodo deposition~\cite{Li2025AEbagging}, whose record metadata specify CC BY 4.0; access requests are handled through the Zenodo record. The NCF waveforms and period-indexed dispersion targets are publicly available without gating from SeisDispFusion-NCF revision \hashid{afcd805}~\cite{Yu2026SeisDispFusionNCF} under Apache-2.0. That processed release derives its NCF waveforms from the CC BY 4.0 Mendeley Data record \emph{Ambient noise cross-correlations in the North China Craton} (\url{https://doi.org/10.17632/m9ry8nbfwj.1}); both releases should be cited. Locked manifests in the reproducibility package record the exact split membership and row order, and immutable hashes permit byte-level verification against these cited releases. Derived predictions, aggregate results and panel-level figure source data are publicly included in the GPL-3.0 reproducibility package and supplied as \emph{Source Data}. The cited raw datasets are not duplicated in that package.

\section*{Code availability}
The reproducibility archive is publicly available on Hugging Face at immutable revision \hashid{784e843152d2335bf57acdbff65d4b8c61688a6d} (\url{https://huggingface.co/cangyeone/elastic-wave-structure-emerge/tree/784e843152d2335bf57acdbff65d4b8c61688a6d})~\cite{Yu2026ElasticWaveArchive}. It contains analysis and figure-generation code, model weights, selected checkpoints, run configurations, derived predictions, checksums and environment metadata. The primary results and the robustness controls used here are retained under \path{reproducibility/primary} and \path{reproducibility/supplementary/nc_robustness}. The robustness materials include affine heads, derived relational matrices, displaced-window predictions, uncertainty resamples and an executable check that regenerates Supplementary Tables~16--20. The repository DOI \url{https://doi.org/10.57967/hf/10158} identifies an earlier archived revision; the immutable Git link above identifies the release cited here. Cited raw datasets and waveform-feature caches are excluded. The original SeismicXM and PNSN implementations are at \url{https://github.com/cangyeone/seismicxm} and \url{https://github.com/cangyeone/pnsn}.

\section*{Acknowledgements}

The authors thank the creators of the receiver-function quality-control data and the developers of PNSN, SeismicXM, PhaseNet and EQTransformer for making research artifacts available. A generative-AI assistant supported literature organization, language editing and reproducibility scripting. The authors verified the analyses, citations and manuscript and remain responsible for the work. No external funding was declared.

\section*{Competing interests}

The authors declare no competing interests.

\section*{Ethics statement}

This study used non-human seismic and laboratory waveform data and involved no human participants, personal data, animals or field intervention.

\bibliographystyle{unsrtnat}
\bibliography{references}

\clearpage
\suppressfloats[t]
\setcounter{figure}{0}
\setcounter{table}{0}
\setcounter{equation}{0}
\setcounter{section}{0}
\renewcommand{\theHfigure}{S.\arabic{figure}}
\renewcommand{\theHtable}{S.\arabic{table}}
\renewcommand{\theHequation}{S.\arabic{equation}}
\renewcommand{\theHsection}{S.\arabic{section}}
\section*{Supplementary Information}
\captionsetup[table]{name=Supplementary Table}
\captionsetup[figure]{name=Supplementary Figure}
\graphicspath{{Figures/Supplementary/}{Figures/}}

\section*{Supplementary Note 1: Convergent evidence framework}

The primary convergence estimand compares PNSN and SeismicXM sample geometry after referencing each released checkpoint to complete random instances of its own topology. Frozen-readout and native-output experiments separately test target accessibility; successful transfer alone does not establish internal agreement. The expanded controls in Supplementary Note~9 distinguish reproducible AE alignment from calibration-dependent intervention responses, and RF accessibility from a readout-independent pretraining advantage. End-to-end scratch, target-specific networks and classical methods address alternative target-training strategies. Transparent descriptors and period-median or distance baselines contextualize absolute performance.

The locked neural matrix uses one data split. Seed repetitions quantify paired optimization or full random-topology variation and do not create independent populations. Receiver-function (RF) uncertainty is assessed over held-out waveforms, native acoustic-emission (AE) transfer is reported across 12 acquisition channels, and ambient-noise correlation (NCF) analyses retain station-pair and station-network structure. RF station/event identities and AE specimen identities were unavailable, defining their estimands as waveform-unique interpolation and later-acquisition transfer.

\section*{Supplementary Note 2: Data provenance}

RF byte hashes removed 801 repeated rows before splitting. The additional 14,398-waveform ``small earthquake'' array in the source repository has no row-level labels or station/event mapping, and 2,300 float32 waveform hashes overlap the labelled array; it was not used as an external test. AE hashes removed 358 repeats across the two files, with no cross-split waveform hash. The NCF manifest repaired endpoint ordering using the source dispersion filename, removed 19 coordinate-ambiguous station codes, and discarded every cross-boundary pair. Pair, station, exact-waveform and complete-target-curve overlaps between development and station-disjoint test were all zero.

\begin{table}[H]
\centering
\caption{\textbf{Locked target datasets and estimands.}}
\begin{adjustbox}{max width=\linewidth}
\begin{tabular}{lllll}
\toprule
Target & Training & Validation & Test & Strongest supported scope \\
\midrule
RF quality & 3,257 & 699 & 699 & waveform-unique interpolation \\
AE detection/pick & 8,728 records & 970 records & 2,073 records$\times2$ crops & later-acquisition transfer \\
NCF velocity/mask & 15,755 & 1,769 & 3,025 & 132-station geographic holdout \\
\bottomrule
\end{tabular}
\end{adjustbox}
\label{si:tab:data}
\end{table}

\begin{table}[H]
\centering
\caption{\textbf{Audited source-model and frozen-representation definitions.} Counts sum each learned tensor once and exclude non-parameter buffers. Retained-path counts exclude source decoders and all target-task readouts.}
\resizebox{\textwidth}{!}{%
\begin{adjustbox}{max width=\linewidth}
\begin{tabular}{lrrrll}
\toprule
Model & Complete model & Retained path & Channels & Source role & Checkpoint SHA-256 \\
\midrule
PNSN v3 & 450,485 & 345,600 & 96 & regional phase picker & \hashid{9f626e5fff4e9390c88e43c2f6230802496163b5d6eefee05e1b6ac7ccebf9e8} \\
SeismicXM & 51,895,545 & 41,332,096 & 1,024 & multi-task waveform model & \hashid{671d02d677c25c3d075963889602299ec71f52c724470f2fa85bb28035fe1528} \\
\bottomrule
\end{tabular}
\end{adjustbox}}
\label{si:tab:encoders}
\end{table}

\section*{Supplementary Note 3: Formal run-level results}

\begin{table}[H]
\centering
\caption{\textbf{Locked transfer results.} RF values are mean $\pm$ sample s.d. over ten paired readouts. AE uses the unchanged released five-class pointwise phase output (one deterministic pretrained checkpoint and five complete random topologies) with no AE-trained neural weights. NCF rows are pooled affine probes over ten paired seeds.}
\begin{adjustbox}{max width=\linewidth}
\begin{tabular}{lllrrr}
\toprule
Task & Encoder & Metric & Pretrained & Random & Favorable gain \\
\midrule
RF & PNSN & balanced accuracy & $0.8113\pm0.0115$ & $0.5353\pm0.0585$ & $+0.2760$ \\
RF & SeismicXM & balanced accuracy & $0.8552\pm0.0100$ & $0.7836\pm0.0224$ & $+0.0716$ \\
AE & PNSN & balanced accuracy & $0.7098$ & $0.5087\pm0.0194$ & $+0.2012$ \\
AE & SeismicXM & balanced accuracy & $0.8254$ & $0.5370\pm0.0456$ & $+0.2883$ \\
AE & PNSN & pick median error (samples) & $33$ & $5117.0\pm1.9$ & $+5084.0$ \\
AE & SeismicXM & pick median error (samples) & $32$ & $5119.4\pm0.5$ & $+5087.4$ \\
NCF & PNSN & velocity MAE (\kms) & $0.0598\pm0.0025$ & $0.0538\pm0.0003$ & $-0.0059$ \\
NCF & SeismicXM & velocity MAE (\kms) & $0.0835\pm0.0077$ & $0.0957\pm0.0188$ & $+0.0122$ \\
NCF & PNSN & mask balanced accuracy & $0.9031\pm0.0014$ & $0.8654\pm0.0026$ & $+0.0377$ \\
NCF & SeismicXM & mask balanced accuracy & $0.9119\pm0.0031$ & $0.8639\pm0.0117$ & $+0.0480$ \\
\bottomrule
\end{tabular}
\end{adjustbox}
\label{si:tab:primary}
\end{table}

\begin{table}[H]
\centering
\caption{\textbf{Native AE localization coverage.} Fractions use all 2,073 positive test windows or the subset whose detection score exceeded the validation-locked scalar threshold. Random values pool five complete random topologies.}
\begin{adjustbox}{max width=\linewidth}
\begin{tabular}{lllrrrrr}
\toprule
Pathway & Population & Median & $\leq25$ & $\leq50$ & $\leq100$ & $\leq200$ \\
\midrule
PNSN pretrained & all positives & 33 & 0.437 & 0.548 & 0.645 & 0.731 \\
PNSN pretrained & detected positives & 19 & 0.569 & 0.719 & 0.847 & 0.943 \\
SeismicXM pretrained & all positives & 32 & 0.396 & 0.599 & 0.756 & 0.854 \\
SeismicXM pretrained & detected positives & 32 & 0.418 & 0.626 & 0.785 & 0.877 \\
PNSN random & all positives & 5,116 & 0.000 & 0.000 & 0.000 & 0.000 \\
SeismicXM random & all positives & 5,119 & 0.000 & 0.000 & 0.000 & 0.000 \\
\bottomrule
\end{tabular}
\end{adjustbox}
\label{si:tab:ae-coverage}
\end{table}

\begin{table}[H]
\centering
\caption{\textbf{Native AE detection across acquisition channels.} Balanced accuracy uses each model's single global validation-locked threshold; no channel-specific refitting was performed.}
\begin{adjustbox}{max width=\linewidth}
\begin{tabular}{lrr@{\hspace{1.5em}}lrr}
\toprule
Channel & PNSN & SeismicXM & Channel & PNSN & SeismicXM \\
\midrule
1 & 0.659 & 0.892 & 7  & 0.640 & 0.807 \\
2 & 0.897 & 0.840 & 8  & 0.732 & 0.837 \\
3 & 0.699 & 0.861 & 9  & 0.645 & 0.792 \\
4 & 0.831 & 0.862 & 10 & 0.595 & 0.821 \\
5 & 0.725 & 0.787 & 11 & 0.606 & 0.789 \\
6 & 0.764 & 0.757 & 12 & 0.644 & 0.760 \\
\bottomrule
\end{tabular}
\end{adjustbox}
\label{si:tab:ae-channel}
\end{table}

\begin{table}[H]
\centering
\caption{\textbf{NCF readout mechanism probes.} Temporal-linear values use five pairs and single temporal-MLP values use ten. Inference ensembles average independently trained members; $K\in\{5,10\}$ was selected separately on validation MAE before test evaluation. The period--distance residual ensemble averages three fixed members for each initialization.}
\begin{adjustbox}{max width=\linewidth}
\begin{tabular}{llrrr}
\toprule
Readout & Encoder & Pretrained MAE & Random MAE & Favorable gain (\kms) \\
\midrule
Pooled affine & PNSN & 0.0598 & 0.0538 & $-0.0059$ \\
Pooled affine & SeismicXM & 0.0835 & 0.0957 & $+0.0122$ \\
Temporal affine & PNSN & 0.0694 & 0.0511 & $-0.0183$ \\
Temporal affine & SeismicXM & 0.0623 & 0.0566 & $-0.0057$ \\
Temporal MLP & PNSN & 0.0589 & 0.0519 & $-0.0070$ \\
Temporal MLP & SeismicXM & 0.0541 & 0.0568 & $+0.0027$ \\
Inference ensemble & PNSN & 0.0570 ($K=10$) & 0.0520 ($K=5$) & $-0.0051$ \\
Inference ensemble & SeismicXM & 0.0538 ($K=5$) & 0.0561 ($K=10$) & $+0.0024$ \\
Period--distance residual ensemble & SeismicXM & 0.0538 ($K=3$) & 0.0562 ($K=3$) & $+0.0024$ \\
\bottomrule
\end{tabular}
\end{adjustbox}
\label{si:tab:readouts}
\end{table}

\begin{table}[H]
\centering
\caption{\textbf{Population and block bootstrap.} Ten thousand percentile resamples; positive gains favor pretraining.}
\resizebox{\textwidth}{!}{%
\begin{adjustbox}{max width=\linewidth}
\begin{tabular}{llllrr}
\toprule
Task & Encoder & Resampling unit & Metric & Gain & 95\% interval \\
\midrule
RF & PNSN & waveform within class & balanced accuracy & 0.2760 & [0.2399, 0.3103] \\
RF & SeismicXM & waveform within class & balanced accuracy & 0.0714 & [0.0442, 0.0999] \\
NCF & PNSN & station pair & velocity MAE & $-0.0070$ & [$-0.0075$, $-0.0066$] \\
NCF & SeismicXM & station pair & velocity MAE & 0.0027 & [0.0022, 0.0032] \\
NCF & PNSN & station-network node & velocity MAE & $-0.0070$ & [$-0.0089$, $-0.0051$] \\
NCF & SeismicXM & station-network node & velocity MAE & 0.0027 & [0.0005, 0.0050] \\
NCF ensemble & SeismicXM & station-network node & random MAE $-$ pretrained MAE & 0.00237 & [0.00004, 0.00484] \\
NCF ensemble & SeismicXM & station-network node & period-median MAE $-$ pretrained MAE & $-0.00313$ & [$-0.00579$, $-0.00037$] \\
NCF residual & SeismicXM & station-network node & random residual MAE $-$ pretrained residual MAE & 0.00243 & [$-0.00014$, 0.00524] \\
NCF residual & SeismicXM & station-network node & period--distance baseline MAE $-$ pretrained residual MAE & 0.00383 & [0.00123, 0.00665] \\
\bottomrule
\end{tabular}
\end{adjustbox}}
\label{si:tab:bootstrap}
\end{table}

\section*{Supplementary Note 4: Label efficiency and transparent baselines}

\begin{table}[H]
\centering
\caption{\textbf{RF standardized logistic analysis.} Mean test balanced accuracy over five label subsamples or matched random encoders.}
\begin{adjustbox}{max width=\linewidth}
\begin{tabular}{llrrrrr}
\toprule
Representation & Initialization & 1\% & 5\% & 10\% & 25\% & 100\% \\
\midrule
Handcrafted & descriptors & 0.8097 & 0.8151 & 0.8085 & 0.8216 & 0.8247 \\
PNSN & pretrained & 0.7445 & 0.7800 & 0.8104 & 0.8058 & 0.8159 \\
PNSN & random & 0.7369 & 0.7619 & 0.7872 & 0.8019 & 0.8163 \\
SeismicXM & pretrained & 0.7368 & 0.7936 & 0.8261 & 0.8440 & 0.8433 \\
SeismicXM & random & 0.6729 & 0.7382 & 0.7750 & 0.8105 & 0.8533 \\
\bottomrule
\end{tabular}
\end{adjustbox}
\label{si:tab:labels}
\end{table}

SeismicXM exceeded its matched random topology at 1--25\% of RF labels, but not at the full label budget; PNSN differences were smaller and likewise did not remain positive at 100\%. These subsampling results therefore diagnose low- and intermediate-budget accessibility, not a monotonic or universal sample-efficiency advantage.

\begin{table}[H]
\centering
\caption{\textbf{Locked-region transparent NCF baselines.} Hyperparameters were selected on validation and test was evaluated once.}
\begin{adjustbox}{max width=\linewidth}
\begin{tabular}{lrrr}
\toprule
Baseline & Selected degree & Selected $\alpha$ & Test MAE (\kms) \\
\midrule
Training period median & -- & -- & 0.05063 \\
Distance-only polynomial ridge & 4 & 0.01 & 0.05762 \\
Waveform-descriptor ridge & -- & 100 & 0.06149 \\
Waveform + distance ridge & 1 & 100 & 0.06073 \\
Waveform + distance, log-travel-time target & 1 & 0.1 & 0.17478 \\
\bottomrule
\end{tabular}
\end{adjustbox}
\label{si:tab:ncfbaselines}
\end{table}

\begin{table}[H]
\centering
\caption{\textbf{Five-cluster geographic analysis for deterministic NCF baselines.} Cluster 1, the largest within-cluster population, was the locked neural test region. Neural encoders were not retrained for clusters 0, 2, 3 or 4.}
\begin{adjustbox}{max width=\linewidth}
\begin{tabular}{rrrrrr}
\toprule
Held-out cluster & Test stations & Test entries & Mean distance (km) & Period median MAE & Distance-only MAE \\
\midrule
0 & 27 & 173 & 288.0 & 0.0821 & 0.0902 \\
1 & 132 & 3,025 & 276.7 & 0.0506 & 0.0576 \\
2 & 102 & 2,230 & 267.8 & 0.0869 & 0.0872 \\
3 & 81 & 887 & 246.9 & 0.0638 & 0.0714 \\
4 & 33 & 238 & 275.3 & 0.0554 & 0.0578 \\
\bottomrule
\end{tabular}
\end{adjustbox}
\label{si:tab:geographic}
\end{table}

\section*{Supplementary Note 5: NCF geometry and physical interpretation}

NCF velocity labels were copied from the original dispersion files onto 2--50-s period bins; they were not recomputed from distance at HDF5 construction. Coordinates and distance were excluded from neural inputs. Nevertheless, the fixed zero-lag origin made distance recoverable from the waveform: Spearman correlations between distance and energy temporal centroid were 0.907, 0.895 and 0.883 in train, validation and test. Correlations with envelope-peak lag were 0.730, 0.727 and 0.675. The manuscript therefore describes geometry as implicit rather than absent.

The validation-selected SeismicXM inference ensemble reached 0.05376~\kms MAE, compared with 0.05613~\kms for its random ensemble and 0.05063~\kms for the training-set period-wise median. Validation selected five pretrained and ten random members. The favorable pretrained--random gain was 0.00237~\kms; a 10,000-resample station-network-node bootstrap gave [0.00004, 0.00484]~\kms and a positive fraction of 0.977. The period-median-minus-pretrained contrast was instead $-0.00313$~\kms, with [$-0.00579$, $-0.00037$]~\kms and a positive fraction of 0.015, so the waveform ensemble did not outperform this transparent baseline. The mean single-member pretrained MAE was 0.05414~\kms, and the member selected by validation reached 0.05401~\kms on test.

The explicit training-only period--distance baseline used degree 4 and ridge $\alpha=0.01$, selected on validation, and reached 0.05762~\kms on test. Matched fixed three-member residual ensembles reconstructed $\hat v(T,d,x)=v_{\mathrm{base}}(T,d)+\Delta v(T,x)$ from frozen pretrained or random SeismicXM sequences. The pretrained ensemble reached 0.05378~\kms, compared with 0.05622~\kms for matched random topology. The random-residual-minus-pretrained-residual gain was 0.00243~\kms; a 10,000-resample paired station-network-node bootstrap gave [$-0.00014$, 0.00524]~\kms with a positive fraction of 0.968. The period--distance-baseline-minus-pretrained-residual gain was 0.00383~\kms, with [0.00123, 0.00665]~\kms and a positive fraction of 0.999. Thus the waveform-conditioned residual improves robustly on the explicit distance baseline, whereas its contrast with random SeismicXM topology is directional but not interval-separated. These diagnostics use station-disjoint paths; crustal inversion and structural-resolution claims are outside this experiment.

\section*{Supplementary Note 6: Direct convergence controls and interventions}

The primary convergence estimand subtracts the average sample geometry of five complete random instances from each pretrained topology before cross-model alignment. Each random control instead subtracts the leave-one-out mean of the other four random instances for both RDMs and kernels. On the positive AE subset, unmodified final-layer total geometry gave standard RSA 0.702 and conventional linear CKA 0.803. The random-subtracted signed statistic is learning-delta kernel alignment, not CKA. Final-layer pretrained--pretrained learning-delta RSA was 0.445 and learning-delta kernel alignment was 0.386. Intermediate-layer learning-delta RSA was 0.450; rank-RDM and signed kernel relations capture complementary organizations.

Waveform-bootstrap 95\% intervals for final AE pretrained--pretrained alignment were [0.332, 0.552] for learning-delta RSA and [0.169, 0.571] for learning-delta kernel alignment. Repeated balanced 75\% subsampling gave [0.378, 0.509] and [0.254, 0.500]. None of 1,000 independent correspondence permutations reached either observed alignment.

\begin{table}[H]
\centering
\caption{\textbf{AE positive-subset learning-delta interventions.} Values compare final-layer PNSN and SeismicXM geometry after subtracting the intervention-matched mean random geometry for each topology.}
\begin{adjustbox}{max width=\linewidth}
\begin{tabular}{lrr}
\toprule
Intervention & Learning-delta RSA $\rho$ & Learning-delta kernel alignment \\
\midrule
Original & 0.445 & 0.386 \\
Phase randomized & 0.157 & 0.171 \\
Polarity reversed & 0.413 & 0.351 \\
Onset neighbourhood removed & 0.300 & 0.082 \\
\bottomrule
\end{tabular}
\end{adjustbox}
\label{si:tab:convergence-interventions}
\end{table}

Phase randomization retained the amplitude spectrum. Onset removal replaced the central $\pm512$ samples by a boundary-matched linear path, avoiding artificial edges from hard zeroing. The reductions describe the original 64-positive-record subset; they do not by themselves establish a calibration-independent mechanism. The larger sample and both training calibrations are reported in Supplementary Note~9 and main Fig.~3. Polarity reversal preserved AE alignment at both final and intermediate depth (intermediate learning-delta RSA 0.484). Native-output polarity reversal likewise retained PNSN/SeismicXM balanced accuracies of 0.696/0.825 and median pick errors of 38/32 samples under the original locked thresholds. These results support robustness to global sign reversal within the tested protocols.

\begin{figure}[H]
\centering
\includegraphics[width=0.72\textwidth]{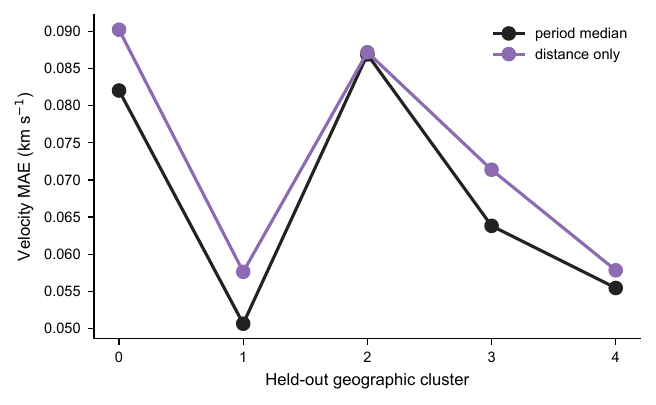}
\caption{\textbf{Geographic variability of deterministic NCF baselines.} Period-median and validation-selected distance-only test MAE for each held-out coordinate cluster.}
\label{si:fig:geographic}
\end{figure}

\section*{Supplementary Note 7: Generality across additional phase-model families}

We tested whether native earthquake-to-AE transfer was restricted to the PNSN--SeismicXM pair by evaluating the original PhaseNet and conservative original EQTransformer weights distributed through SeisBench~\cite{ZhuBeroza2019,MousaviEtAl2020EQT,WoollamEtAl2022}. Their functional transfer is included in main Fig.~2, but they do not redefine the primary convergence estimand. They provide an out-of-pair generality test of operational arrival structure; a separate common-input matrix below tests, rather than assumes, whether all four architectures share source-induced geometry.

The additional test retained the locked AE validation and test records. A common 6,000-sample crop was used because this is EQTransformer's fixed input length and is accepted without architectural modification by fully convolutional PhaseNet. The positive annotation was placed at sample 3,000; the paired negative crop ended 100 samples before the annotation. Each scalar AE trace was repeated into three identical components. PhaseNet inputs were demeaned and divided by per-component standard deviation. EQTransformer inputs were demeaned, divided by the joint standard deviation and tapered over six samples at each edge. No frequency filter was imposed because AE sampling-rate metadata were unavailable; the sample-domain protocol was therefore retained. The central 5,000 outputs were scored after a common 500-sample edge exclusion on each side.

PhaseNet detection used the maximum released P probability and picking used its argmax. EQTransformer detection used the maximum released detection probability and picking used the argmax of its released P output. A single scalar threshold for each model and initialization maximized validation balanced accuracy with F1 and proximity-to-0.5 tie breaking, then remained locked on test. No AE-labelled neural parameter, calibration layer or target decoder was fitted. Five complete random instances of each exact topology supplied matched controls. PhaseNet and EQTransformer contain 268,443 and 376,935 parameters, respectively. Their cached original-weight SHA-256 values were \hashid{184dd85b896eebf8d5967ee15a722a477fa2e4ce479a3b113ecbcb8a6419113f} and \hashid{c363c59dfb4562ba3b1356fb03dc9c776bbcfe3384b5db22e5ee3689de0a4df9}.

Both additional pretrained models transferred well above matched random topology (Supplementary Table~\ref{si:tab:additional-phase-models}). PhaseNet reached test balanced accuracy 0.9033, compared with a five-instance random mean of $0.5078\pm0.0106$. Its favorable gain was 0.3955, with a 10,000-resample paired-record 95\% interval of [0.3865, 0.4044]. EQTransformer reached 0.9267, compared with a random mean of $0.5000\pm0.0000$, giving a gain of 0.4267 [0.4187, 0.4342]. All bootstrap replicates were positive for both contrasts.

The released P outputs also localized AE onsets. PhaseNet's median absolute error was 67 samples, compared with a random-topology mean of $748.8\pm54.9$; the random-minus-pretrained median-error gain was 681.8 [655.8, 704.0] samples. EQTransformer's median error was 145 samples, compared with $2437.0\pm78.7$ for random topology, giving 2292 [2280, 2296] samples. The complete error distributions retained long tails, especially for EQTransformer, so detection and localization are reported separately. PhaseNet localized 28.2\%, 41.2\%, 62.0\% and 81.5\% of all positive windows within $\pm25$, $\pm50$, $\pm100$ and $\pm200$ samples; the corresponding EQTransformer fractions were 15.6\%, 24.9\%, 39.7\% and 54.5\%.

Performance was not confined to one acquisition channel. Under each model's single global validation-locked threshold, PhaseNet channel-wise balanced accuracy ranged from 0.838 to 0.938 and EQTransformer ranged from 0.887 to 0.968 across all 12 channels (Supplementary Table~\ref{si:tab:additional-phase-channels}). These out-of-pair results broaden the evidence that earthquake-trained phase and detection objectives recover reusable arrival structure across the earthquake-to-laboratory scale change. Functional transfer, however, does not by itself establish identical internal geometry.

\begin{table}[H]
\centering
\caption{\textbf{Native-output AE transfer for two additional pretrained phase-model families.} Random entries are mean $\pm$ sample s.d. across five complete random topologies. Balanced-accuracy gain is pretrained minus random; pick gain is random median error minus pretrained median error. Intervals are 10,000-resample paired-record percentile intervals.}
\resizebox{\textwidth}{!}{%
\begin{adjustbox}{max width=\linewidth}
\begin{tabular}{lrrrrrr}
\toprule
Model & Pretrained BA & Random BA & BA gain [95\% interval] & Pretrained median error & Random median error & Pick gain [95\% interval] \\
\midrule
PhaseNet & 0.9033 & $0.5078\pm0.0106$ & 0.3955 [0.3865, 0.4044] & 67 & $748.8\pm54.9$ & 681.8 [655.8, 704.0] \\
EQTransformer & 0.9267 & $0.5000\pm0.0000$ & 0.4267 [0.4187, 0.4342] & 145 & $2437.0\pm78.7$ & 2292 [2280, 2296] \\
\bottomrule
\end{tabular}
\end{adjustbox}}
\label{si:tab:additional-phase-models}
\end{table}

\begin{table}[H]
\centering
\caption{\textbf{Additional pretrained-model AE detection across acquisition channels.} Balanced accuracy uses one global validation-locked threshold per model; no channel-specific threshold was fitted.}
\begin{adjustbox}{max width=\linewidth}
\begin{tabular}{lrr@{\hspace{1.5em}}lrr}
\toprule
Channel & PhaseNet & EQTransformer & Channel & PhaseNet & EQTransformer \\
\midrule
1 & 0.938 & 0.950 & 7  & 0.860 & 0.895 \\
2 & 0.854 & 0.892 & 8  & 0.923 & 0.947 \\
3 & 0.910 & 0.968 & 9  & 0.889 & 0.942 \\
4 & 0.920 & 0.937 & 10 & 0.934 & 0.887 \\
5 & 0.884 & 0.932 & 11 & 0.914 & 0.906 \\
6 & 0.838 & 0.905 & 12 & 0.909 & 0.904 \\
\bottomrule
\end{tabular}
\end{adjustbox}
\label{si:tab:additional-phase-channels}
\end{table}

To test internal geometry directly, we evaluated all four pretrained checkpoints and five complete random topologies per model on the identical locked AE subset and common 6,000-sample crop. PNSN used its final bidirectional recurrent sequence and SeismicXM its final waveform-token sequence, each summarized by a 16-bin adaptive mean--maximum grid. PhaseNet used its flattened encoder bottleneck, whereas EQTransformer used the centre position of its shared post-attention sequence. Each representation was standardized from its own training subset. Learning-delta RDMs and centered kernels subtracted the corresponding model's mean random geometry before pairwise comparison.

\begin{figure}[H]
\centering
\includegraphics[width=\textwidth]{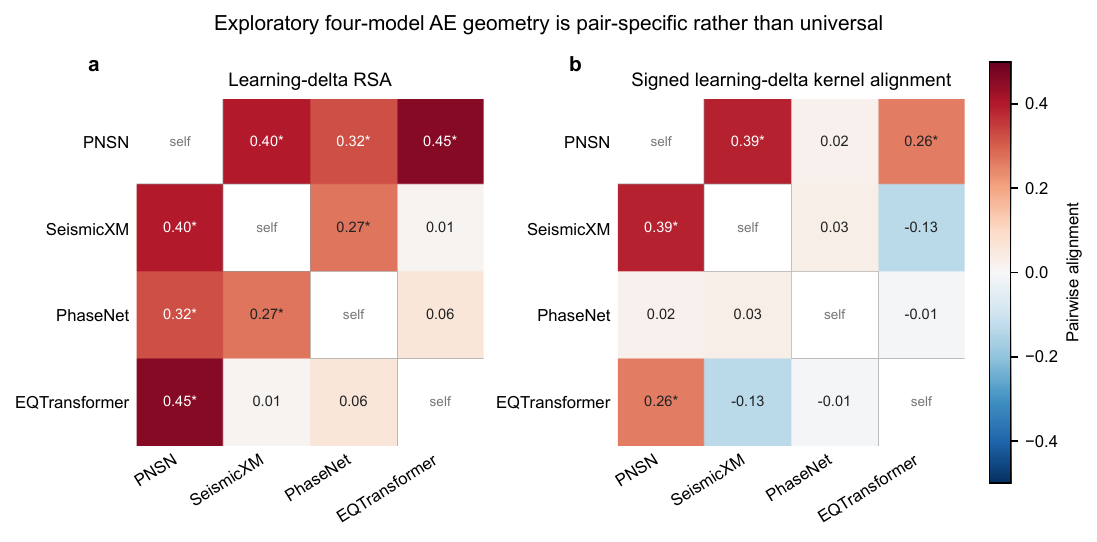}
\caption{\textbf{Exploratory four-model AE learning-delta convergence.} \textbf{a}, Pairwise RSA of source-induced cosine-RDM geometry. \textbf{b}, Pairwise signed alignment of source-induced centered kernels. All models received the same 6,000-sample crops and were referenced to five complete matched-random topologies. Values are evaluated on 64 positive held-out AE windows; asterisks mark one-sided correspondence-permutation $P<0.05$ from 1,000 permutations. The nonuniform matrix does not support an all-to-all common latent geometry.}
\label{si:fig:four-model-convergence}
\end{figure}

The matrix was nonuniform (Supplementary Fig.~\ref{si:fig:four-model-convergence}; Supplementary Table~\ref{si:tab:four-model-convergence}). PNSN--SeismicXM remained positive under both metrics on the common crop, as did PNSN--EQTransformer. PNSN--PhaseNet and SeismicXM--PhaseNet were positive under RSA but not under the signed kernel test. SeismicXM--EQTransformer and PhaseNet--EQTransformer did not separate from correspondence permutations. Thus, the four families support broad native-output arrival transfer, but the present interfaces do not support an all-to-all common latent geometry. The stronger convergence claim remains conditional on the primary PNSN--SeismicXM routes; the exploratory matrix defines a boundary rather than a failed universality requirement.

\begin{table}[H]
\centering
\caption{\textbf{Pairwise four-model AE learning-delta alignment.} Intervals are 1,000-resample waveform-bootstrap percentile intervals. $P$ values are one-sided against 1,000 independent correspondence permutations.}
\resizebox{\textwidth}{!}{%
\begin{adjustbox}{max width=\linewidth}
\begin{tabular}{lrrrr}
\toprule
Model pair & RSA [95\% interval] & Permutation $P$ & Kernel alignment [95\% interval] & Permutation $P$ \\
\midrule
PNSN--SeismicXM & 0.397 [0.255, 0.512] & 0.001 & 0.385 [0.227, 0.525] & 0.001 \\
PNSN--PhaseNet & 0.324 [0.188, 0.439] & 0.001 & 0.022 [0.008, 0.062] & 0.709 \\
PNSN--EQTransformer & 0.454 [0.311, 0.566] & 0.001 & 0.257 [0.105, 0.410] & 0.001 \\
SeismicXM--PhaseNet & 0.266 [0.112, 0.408] & 0.001 & 0.028 [$-0.015$, 0.143] & 0.318 \\
SeismicXM--EQTransformer & 0.015 [$-0.112$, 0.145] & 0.391 & $-0.135$ [$-0.268$, 0.025] & 1.000 \\
PhaseNet--EQTransformer & 0.061 [$-0.101$, 0.205] & 0.104 & $-0.007$ [$-0.058$, 0.020] & 0.501 \\
\bottomrule
\end{tabular}
\end{adjustbox}}
\label{si:tab:four-model-convergence}
\end{table}

We additionally tested the architecture-native frozen representations proposed for NCF dispersion, without changing the main convergence estimand. PhaseNet's 128-channel, six-position encoder bottleneck was flattened to a 768-dimensional vector. EQTransformer requires 6,000 input samples, so each 1,501-sample NCF trace was symmetrically zero padded (2,249 samples left and 2,250 right), normalized over the padded three-component window and mapped to the centre position (index 23) of its 16-channel, 47-position shared post-attention sequence. These choices were fixed before testing. Each representation fed one affine map producing 49 velocity values and 49 mask logits. The head and loss matched the locked NCF linear-readout protocol; model selection used validation velocity MAE, and the held-out station-disjoint test set was evaluated once. Five paired seeds compared each released checkpoint with a complete random instance of the same topology. Head capacity was therefore matched within each model (75,362 parameters for PhaseNet and 1,666 for EQTransformer), although it was deliberately not equated across architectures.

Neither architecture-native interface provided positive pretrained-to-random NCF transfer (Supplementary Table~\ref{si:tab:additional-phase-ncf}). PhaseNet's flattened bottleneck reached $0.05778\pm0.00266$~km~s$^{-1}$ test MAE, compared with $0.05473\pm0.00072$ for random topology. The paired-seed random-minus-pretrained gain was $-0.00305\pm0.00261$~km~s$^{-1}$ and favored pretraining in only one of five seeds. For five-seed ensemble predictions, the gain was $-0.00046$~km~s$^{-1}$; its 10,000-resample station-network-node interval, [$-0.00280$, $0.00191$], included zero. EQTransformer's centre-position representation reached $0.07284\pm0.00098$~km~s$^{-1}$, versus $0.05332\pm0.00050$ for random topology. Its paired-seed gain was $-0.01952\pm0.00082$~km~s$^{-1}$ and was unfavorable in all five seeds; the ensemble station-network-node gain was $-0.01894$ [$-0.02413$, $-0.01395$]~km~s$^{-1}$.

These controls separate arrival-transfer generality from universal dispersion composability. PhaseNet and EQTransformer retain strong native earthquake-to-AE arrival transfer, but a flattened convolutional bottleneck or one globally contextualized EQTransformer position does not establish a pretrained advantage for NCF dispersion. The result supports treating cross-objective composition as representation- and interface-specific rather than as an automatic consequence of successful phase training.

\begin{table}[H]
\centering
\caption{\textbf{Architecture-native NCF dispersion readouts for the additional phase-model families.} MAE values are mean $\pm$ sample s.d. across five paired head seeds in km~s$^{-1}$. Gain is random MAE minus pretrained MAE, so positive values favor pretraining. The final column compares five-seed ensemble predictions and reports a 10,000-resample station-network-node 95\% interval.}
\resizebox{\textwidth}{!}{%
\begin{adjustbox}{max width=\linewidth}
\begin{tabular}{llrrrr}
\toprule
Model & Frozen representation & Pretrained MAE & Random MAE & Paired-seed gain & Ensemble gain [node 95\% interval] \\
\midrule
PhaseNet & Flattened $128\times6$ bottleneck & $0.05778\pm0.00266$ & $0.05473\pm0.00072$ & $-0.00305\pm0.00261$ & $-0.00046$ [$-0.00280$, $0.00191$] \\
EQTransformer & Centre post-attention position, 16 dimensions & $0.07284\pm0.00098$ & $0.05332\pm0.00050$ & $-0.01952\pm0.00082$ & $-0.01894$ [$-0.02413$, $-0.01395$] \\
\bottomrule
\end{tabular}
\end{adjustbox}}
\label{si:tab:additional-phase-ncf}
\end{table}

\section*{Supplementary Note 8: Executed artifacts}

The deposited archive is rooted at \path{reproducibility/}. It contains 240 complete formal readout runs under \path{primary/runs/formal}, 20 temporal-linear runs under \path{primary/runs/ncf_temporal_probe_v3} and 40 independently optimized temporal-MLP members under \path{primary/runs/ncf_temporal_mlp_probe_v1}. The native AE archive stores 12 complete-model evaluations---two pretrained checkpoints and ten matched random topologies---with pointwise curves, validation/test scores, picks, locked-threshold polarity controls and localization distributions under \path{primary/analysis/native_ae_phase}. Direct convergence stores every initialization condition, layer and intervention, together with bootstrap, repeated-subset and permutation outputs, under \path{primary/analysis/representation_convergence}. Ensemble validation predictions, selected sizes, member diversity and test predictions are stored under \path{primary/analysis/ncf_inference_ensemble}. The fixed three-member period--distance residual runs store histories, selected checkpoints, metrics and full prediction arrays under \path{primary/runs/ncf_distance_residual_ensemble_v1}; their aggregate and paired station-network-node uncertainty outputs are under \path{primary/analysis/ncf_distance_residual_ensemble}. Supplementary PNSN final-state controls are deposited under \path{supplementary/pnsn_final_state}; PhaseNet--EQTransformer AE and architecture-native NCF controls, including all validation/test outputs and 10,000-resample uncertainty products, are under \path{supplementary/phasenet_eqtransformer_ae} and \path{supplementary/phasenet_eqtransformer_ncf}. The exploratory common-input matrices, pairwise estimates, uncertainty resamples and figure source are under \path{supplementary/four_model_ae_convergence}. The release retains all 338 saved downstream checkpoints and a file-level \path{ARTIFACTS.sha256} inventory. All 79 checkpoint-dependent pretrained SeismicXM run records point to \path{models/seismicxm/seismicxm.middle.pt}, whose SHA-256 is \hashid{671d02d677c25c3d075963889602299ec71f52c724470f2fa85bb28035fe1528}. Machine-readable aggregate tables, the derived NCF implicit-geometry summary, checkpoint hashes, figure source data, executed script snapshots and the executable manuscript--artifact consistency gate are deposited together in the reproducibility package~\cite{Yu2026ElasticWaveArchive}; cited raw datasets and waveform-feature caches are not redistributed.

The post-review robustness package under \path{supplementary/nc_robustness} additionally contains all 48 selected RF readouts and their coefficients, held-out probabilities, native displaced-window scores and picks, expanded AE and synthetic relational matrices, paired resamples and run provenance. The five completed analyses retain their official checkpoint hashes and code hashes. \path{scripts/check_robustness_results.py} recomputes the additional statistics and generates Supplementary Tables~16--20; \path{scripts/manuscript_figures/generate_robustness_figures.py} generates the new main Fig.~3 and supplementary control figure. Original primary results are retained rather than overwritten. No cited input waveforms or feature caches accompany these products.

\section*{Supplementary Note 9: Post-review robustness and boundary tests}

These analyses were specified after the original manuscript review and before their execution, except for the explicitly post-hoc balanced-scaler sensitivity described below. They are extensions on the locked splits, not a preregistered independent replication. All favorable, null and unfavorable outcomes are reported. The source checkpoints were unchanged; each random comparison reconstructs the complete topology with seeds 20260826--20260830. The selection seed for the new controls was 20260907. Source-training seed uncertainty remains unmeasured.

\subsection*{RF interface and standardization factorial}

Both models received the same RF demean/maximum-amplitude input normalization, with no additional source-wrapper normalization. Final representations were either mean--maximum pooled or the first temporal position. For each interface, class-balanced logistic regression was fitted with and without training-only feature standardization (\texttt{liblinear}, 10,000 maximum iterations, tolerance $10^{-5}$). Validation balanced accuracy selected $C\in\{0.01,0.1,1,10\}$, with ties preferring the value closest to one on a logarithmic scale. The same locked 3,257/699/699 RF train/validation/test split was used throughout. All four protocols per model appear in Supplementary Table~\ref{si:tab:robust-rf}. Paired uncertainty resamples test waveforms within class after averaging correctness over the five fixed random instances; it does not treat RDM entries or source-training seeds as independent observations.

Standardized random features matched pretrained performance closely enough that none of the standardized gain intervals excluded zero. Thus, the large locked-protocol PNSN advantage in Supplementary Table~3 is not a readout-invariant pretraining effect. The new controlled factorial changes the solver and validation-selected regularization as well as examining scaling, so it diagnoses protocol dependence rather than assigning all differences from Table~3 to scaling alone. Selected affine coefficients and scaler parameters were reconstructed and checked against every stored test probability. An earlier normalization-mismatched pilot was interrupted, retained locally as a failed implementation record, and excluded from the reported completed experiment.

\subsection*{AE annotation-position control}

We selected 512 test and 192 validation records proportionally across the 12 acquisition channels, without replacing records or changing split membership. Each test record supplied a centred positive crop, a positive crop with the annotation sampled uniformly at an integer position from 1,000 to 5,000 inclusive, and the same pretrigger negative crop. Positive crops contained 6,000 samples. After each model's standard normalization, PNSN and SeismicXM received 144 zero-valued samples on the right to obtain a grid-compatible length of 6,144; their complete outputs were verified to have the same length and were scored only on original positions. PhaseNet and EQTransformer retained 6,000 inputs. All models scored positions 500--5,499, without interpolation of native probabilities.

\begin{figure}[H]
\centering
\includegraphics[width=\textwidth]{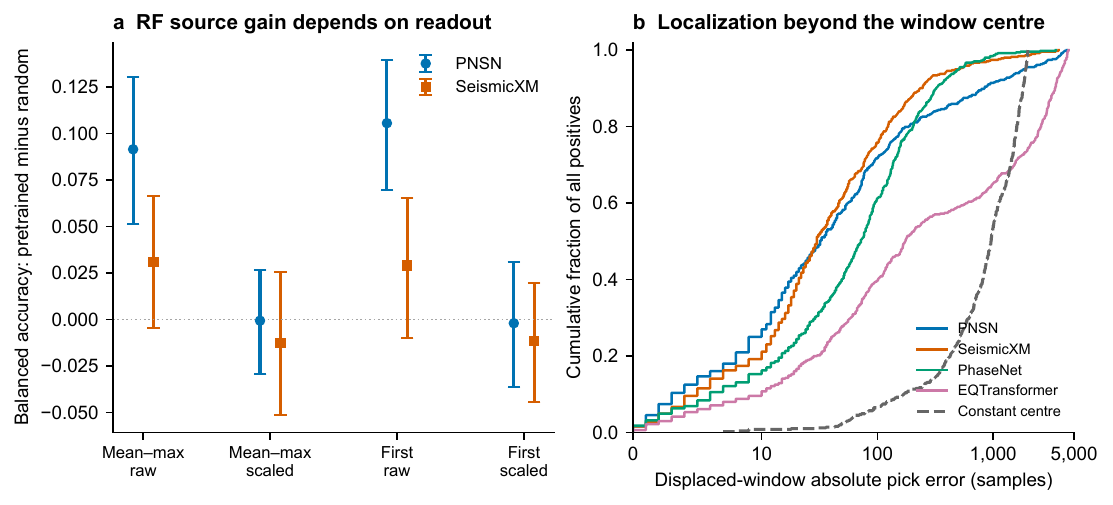}
\caption{\textbf{Readout and arrival-position controls.} \textbf{a}, RF pretrained-minus-random balanced accuracy under all interface/scaling protocols, with paired waveform-bootstrap 95\% intervals. \textbf{b}, Native-output error distributions on displaced AE crops for all positive records, compared with the constant-centre predictor. Zero-error observations are retained on the symmetric-log axis. All thresholds were locked on centred validation data.}
\label{si:fig:rf-ae-robustness}
\end{figure}

Scalar detection thresholds were selected exclusively on centred validation crops and frozen for both test conditions. Localization was evaluated on every positive, not only detections. A constant prediction at sample 3,000 was the explicit position-only reference. Prediction displacement was regressed on annotation displacement; 2,000 paired acquisition-channel bootstrap resamples quantified median error, gain over the constant reference, displacement slope and balanced accuracy. Exact summaries, including long tails, per-channel results and all-positive tolerance rates, are stored with the native predictions. All four models improved median localization over the constant reference after displacement, but EQTransformer had the weakest displacement slope and substantial errors in the distribution tail (Supplementary Fig.~\ref{si:fig:rf-ae-robustness}b). The result does not establish continuous-stream detection or timing in seconds.

\subsection*{Expanded geometry, matched ablation and calibration sensitivity}

The expanded geometry used 512 positive test records and initially 192 positive training records, independently sampled within channel strata. It retained 10,240-sample inputs and the original final-layer 16-bin temporal mean--maximum descriptor. Training-only means and standard deviations were fixed across all test interventions. Phase randomization preserved each Fourier amplitude, including the real-valued DC and Nyquist constraints. Boundary-matched linear replacement removed samples 4,608--5,632 around the onset or 6,656--7,680 in a later window. Both replacements therefore affected exactly 1,025 samples. Polarity reversal changed only the global sign. All random instances received the same transformed inputs as the pretrained models.

After observing the positive-only-calibrated results, we added a separately logged sensitivity test using the exact original 192-example balanced signal/pretrigger training descriptors. This reused identical expanded test features and did not select or retrain a model. Supplementary Tables~\ref{si:tab:robust-geometry} and \ref{si:tab:robust-interventions} report both calibrations; the original calibration is not selected as a new optimum. Geometry bootstrap resampled acquisition channels with replacement, retaining paired sample indices across interventions. Repeated subsets of 64, 128 and 256 of the expanded records are available as descriptive sample-stability outputs, not independent replications.

Original-input alignment remained positive under both calibrations. Phase randomization reduced both measures under balanced calibration but increased signed kernel alignment under positive-only calibration. The onset effect was likewise not invariant to calibration or metric. Late-minus-onset contrasts were positive under both metrics and calibrations, supporting relative sensitivity to waveform position without identifying a unique physical mechanism. Because these are exploratory pointwise intervals, no family-wise significance claim is made. A descriptor control removed rank-RDM associations with 16-bin Hilbert envelopes, 16-bin log-power spectra, and RMS, maximum gradient and post/pre-onset energy ratio. Descriptor standardization used the training subset only. Residual rank association remained positive for AE, but this limited nuisance set cannot exclude every generic signal explanation.

\subsection*{Synthetic scalar-packet boundary}

We generated independent two-arrival signals $x(t)=w(t-t_0)+a\,w(t-t_0-\Delta)+\epsilon(t)$, with $w(u)=[1-2(u/s)^2]\exp[-(u/s)^2]$. Independent uniform draws varied delay $\Delta$ from 200--1,800 samples, width $s$ from 20--80, relative amplitude $a$ from 0.4--1.0, origin $t_0$ from 1,200--1,800 and Gaussian-noise standard deviation from 0.005--0.05. Inputs contained 6,144 samples; 192 training-scaler and 256 test realizations used separate fixed seeds. Fixed-spectrum phase surrogates changed temporal organization without changing the original Fourier amplitudes. The two models and their five random instances used the same inputs and the final 16-bin descriptor. Uncertainty resampled independent realizations 500 times; partial rank association with known delay controlled the distances in the other four generating parameters.

Both learning-delta measures were negative on the original packets and positive on phase surrogates (Supplementary Table~\ref{si:tab:robust-synthetic}). The known-delay associations also differed between the models. These signals do not include vector polarization, mode conversion or an elastic-medium solver, so the outcome neither disproves the measured AE association nor establishes an alternative physical law. It instead limits its extrapolation: the tested seismic representations do not converge on every synthetic arrival-bearing signal family.

\begin{table}[H]
\centering\small
\caption{\textbf{RF readout factorial.} Test balanced accuracy; random means and sample s.d. use five complete topologies. Gain is pretrained minus random mean, with a paired class-stratified waveform-bootstrap 95\% interval (10,000 resamples). All four protocols per model are reported; regularization was selected only on validation.}
\setlength{\tabcolsep}{4pt}
\begin{adjustbox}{max width=\linewidth}
\begin{tabular}{lllll}
\toprule
Model & Interface / scaling & Pretrained & Random & Gain [95\% CI] \\
\midrule
PNSN & Mean--max / raw & 0.8047 & $0.7133\pm0.0613$ & $0.0915$ [0.0512, 0.1305] \\
PNSN & Mean--max / scaled & 0.8159 & $0.8165\pm0.0207$ & $-0.0006$ [-0.0295, 0.0267] \\
PNSN & First / raw & 0.8243 & $0.7187\pm0.0721$ & $0.1055$ [0.0695, 0.1394] \\
PNSN & First / scaled & 0.8132 & $0.8152\pm0.0175$ & $-0.0020$ [-0.0363, 0.0311] \\
SeismicXM & Mean--max / raw & 0.8445 & $0.8137\pm0.0124$ & $0.0308$ [-0.0048, 0.0665] \\
SeismicXM & Mean--max / scaled & 0.8416 & $0.8542\pm0.0073$ & $-0.0126$ [-0.0511, 0.0257] \\
SeismicXM & First / raw & 0.8445 & $0.8155\pm0.0198$ & $0.0290$ [-0.0101, 0.0655] \\
SeismicXM & First / scaled & 0.8662 & $0.8779\pm0.0147$ & $-0.0117$ [-0.0445, 0.0198] \\
\bottomrule
\end{tabular}
\end{adjustbox}
\label{si:tab:robust-rf}
\end{table}

\begin{table}[H]
\centering\small
\caption{\textbf{Native AE localization after random annotation displacement.} Centred/displaced results use the same 512 test records. Errors are in samples, not seconds. Slope regresses the change in prediction on the change in annotation; intervals use 2,000 paired acquisition-channel bootstrap resamples. The constant-centre predictor has median error 0.0 when centred and 959.5 after displacement. Detection thresholds were locked on centred validation crops.}
\setlength{\tabcolsep}{4pt}
\begin{adjustbox}{max width=\linewidth}
\begin{tabular}{llll}
\toprule
Model & BA centred / displaced & Median error centred / displaced & Slope [95\% CI] \\
\midrule
PNSN & 0.7363 / 0.7168 & 45.0 / 32.0 & $0.772$ [0.648, 0.860] \\
SeismicXM & 0.8760 / 0.8672 & 25.0 / 30.0 & $0.904$ [0.826, 0.964] \\
PhaseNet & 0.8926 / 0.8926 & 67.5 / 72.0 & $0.958$ [0.928, 0.986] \\
EQTransformer & 0.9121 / 0.9150 & 172.5 / 174.5 & $0.497$ [0.348, 0.613] \\
\bottomrule
\end{tabular}
\end{adjustbox}
\label{si:tab:robust-ae-position}
\end{table}

\begin{table}[H]
\centering\small
\caption{\textbf{Expanded AE geometry under two fixed training calibrations.} The same 512 positive test records and five random instances per route are used throughout. Balanced scaler retains the original 192-example signal/background calibration; positive scaler uses 192 positive training records. Intervals are acquisition-channel bootstrap 95\% intervals (500 shared resamples). Neither protocol is selected by its test result.}
\setlength{\tabcolsep}{4pt}
\begin{adjustbox}{max width=\linewidth}
\begin{tabular}{llll}
\toprule
Calibration & Input & Learning-delta RSA [95\% CI] & Signed kernel [95\% CI] \\
\midrule
Balanced scaler & Original & $0.414$ [0.316, 0.455] & $0.201$ [0.070, 0.320] \\
Balanced scaler & Polarity reversed & $0.413$ [0.307, 0.460] & $0.207$ [0.069, 0.325] \\
Balanced scaler & Phase randomized & $0.075$ [-0.053, 0.139] & $0.077$ [0.014, 0.123] \\
Balanced scaler & Onset removed & $0.317$ [0.231, 0.375] & $0.124$ [-0.027, 0.237] \\
Balanced scaler & Late window removed & $0.408$ [0.380, 0.433] & $0.307$ [0.267, 0.341] \\
Positive scaler & Original & $0.190$ [0.024, 0.300] & $0.129$ [0.076, 0.197] \\
Positive scaler & Polarity reversed & $0.187$ [0.019, 0.305] & $0.123$ [0.070, 0.192] \\
Positive scaler & Phase randomized & $0.003$ [-0.202, 0.142] & $0.236$ [0.145, 0.318] \\
Positive scaler & Onset removed & $0.183$ [0.063, 0.284] & $0.147$ [0.085, 0.226] \\
Positive scaler & Late window removed & $0.333$ [0.276, 0.391] & $0.274$ [0.249, 0.318] \\
\bottomrule
\end{tabular}
\end{adjustbox}
\label{si:tab:robust-geometry}
\end{table}

\begin{table}[H]
\centering\small
\caption{\textbf{Paired expanded-AE intervention contrasts.} Positive original-minus-intervention values indicate loss of alignment. Late-minus-onset compares equal-length replacements at different waveform positions. Both protocols use identical channel resamples. These are pointwise exploratory 95\% intervals, without family-wise multiplicity adjustment.}
\setlength{\tabcolsep}{4pt}
\begin{adjustbox}{max width=\linewidth}
\begin{tabular}{llll}
\toprule
Calibration & Contrast & RSA difference [95\% CI] & Kernel difference [95\% CI] \\
\midrule
Balanced scaler & Original minus polarity reversed & $0.001$ [-0.010, 0.014] & $-0.006$ [-0.022, 0.010] \\
Balanced scaler & Original minus phase randomized & $0.339$ [0.257, 0.416] & $0.124$ [0.023, 0.254] \\
Balanced scaler & Original minus onset removed & $0.097$ [0.018, 0.143] & $0.077$ [-0.067, 0.233] \\
Balanced scaler & Original minus late window removed & $0.006$ [-0.088, 0.045] & $-0.106$ [-0.247, 0.020] \\
Balanced scaler & Late minus onset & $0.092$ [0.043, 0.182] & $0.183$ [0.082, 0.322] \\
Positive scaler & Original minus polarity reversed & $0.003$ [-0.008, 0.011] & $0.006$ [-0.007, 0.018] \\
Positive scaler & Original minus phase randomized & $0.188$ [-0.004, 0.426] & $-0.107$ [-0.174, -0.005] \\
Positive scaler & Original minus onset removed & $0.007$ [-0.084, 0.057] & $-0.018$ [-0.092, 0.071] \\
Positive scaler & Original minus late window removed & $-0.143$ [-0.268, -0.079] & $-0.145$ [-0.200, -0.080] \\
Positive scaler & Late minus onset & $0.150$ [0.094, 0.242] & $0.127$ [0.075, 0.186] \\
\bottomrule
\end{tabular}
\end{adjustbox}
\label{si:tab:robust-interventions}
\end{table}

\begin{table}[H]
\centering\small
\caption{\textbf{Synthetic two-arrival scalar-packet boundary test.} Learning deltas are referenced to five complete random topologies; 256 independent synthetic test realizations and 192 training-scaler realizations were used. Intervals use 500 realization-bootstrap resamples. Negative alignment is not evidence of a common geometry. The phase surrogates retain the original amplitude spectra; these signals are not a full elastic-wave simulation.}
\setlength{\tabcolsep}{4pt}
\begin{adjustbox}{max width=\linewidth}
\begin{tabular}{lll}
\toprule
Input & Learning-delta RSA [95\% CI] & Signed kernel [95\% CI] \\
\midrule
Original & $-0.404$ [-0.475, -0.333] & $-0.503$ [-0.561, -0.423] \\
Phase randomized & $0.282$ [0.200, 0.346] & $0.141$ [0.110, 0.208] \\
\bottomrule
\end{tabular}
\end{adjustbox}
\label{si:tab:robust-synthetic}
\end{table}

\noindent Residual rank association after envelope, spectrum and elementary onset-descriptor adjustment was 0.339 for positive-calibrated AE and -0.114 for synthetic packets. These are descriptive conditional associations, not causal estimates. Partial association with known synthetic delay, controlling source width, reflection amplitude, origin and noise, was -0.120 / 0.163 (PNSN/SeismicXM) on original packets and 0.020 / 0.005 on phase surrogates.

\begin{figure}[H]
\centering
\includegraphics[width=0.95\textwidth]{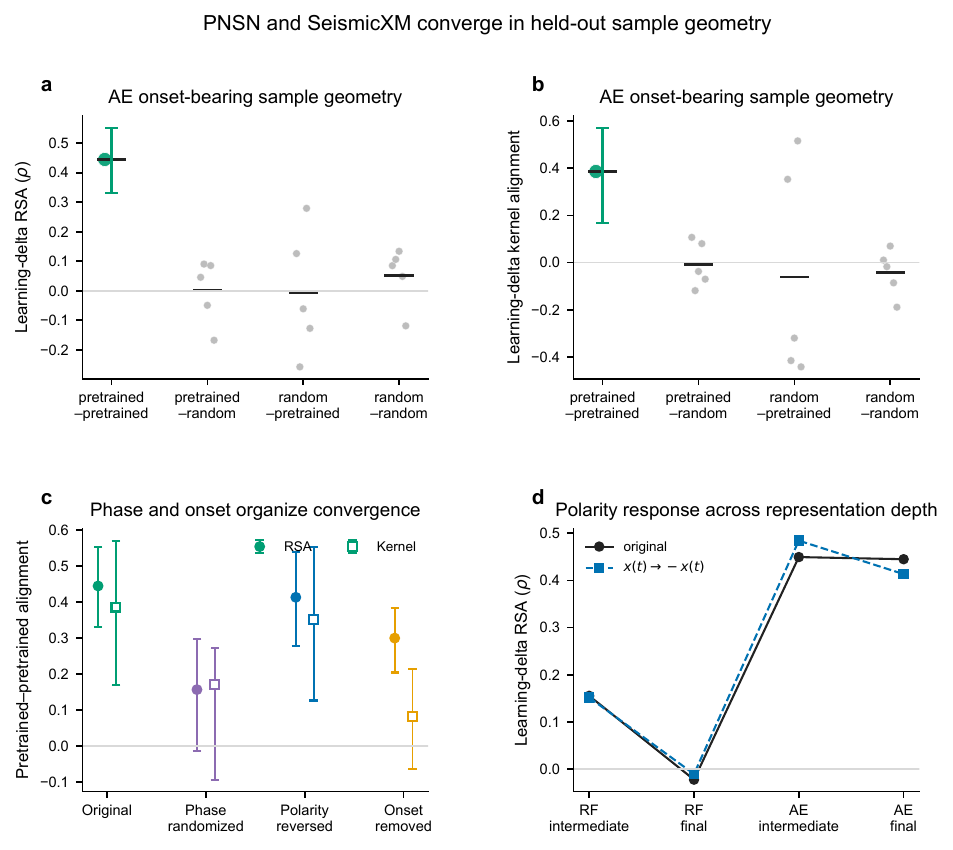}
\caption{\textbf{Original depth and random-control analysis, retained in full.} \textbf{a,b}, Learning-delta RSA and signed kernel alignment for the original AE positive subset relative to complete random topologies; random controls use leave-one-out means. \textbf{c}, Original intervention estimates. \textbf{d}, Intermediate and final RF/AE rank geometry before and after polarity reversal. These original results are supplemented, not replaced, by the expanded sample, paired ablations and calibration controls in main Fig.~3.}
\label{si:fig:original-convergence}
\end{figure}

\end{document}